\documentclass[final,journal]{IEEEtran}

\usepackage{layout, calc}
\usepackage{setspace}
\usepackage{times,amssymb,amsmath,epsfig,amsthm}
\usepackage{citesort}
\usepackage[usenames]{color}
\usepackage{bm}
\newcommand{\myincludegraphics}[2]{\includegraphics#1#2}

\newtheorem{theorem}{Theorem}
   \newtheorem{lemma}{Lemma}
   \newtheorem{corollary}{Corollary}
   
   \newtheorem{remark}{Remark}

\newcommand{\ba}{{\bm{\alpha}}}
\newcommand{\bah}{{\widehat{\bm{\alpha}}}}
\newcommand{\jt}{{J_{\mathsf T}}}
\newcommand{\jr}{{J_{\mathsf R}}}
\newcommand{\hr}{{{\widehat{H}_{2}}}}
\newcommand{\snr}{{{\mathsf{ SNR}}}}
\title{Power-Controlled Feedback and Training for Two-way MIMO Channels}

\author{ Vaneet Aggarwal, and Ashutosh Sabharwal,~\IEEEmembership{Senior Member,~IEEE}  \\
\thanks{V. Aggarwal is with the Department of Electrical Engineering,
Princeton University, Princeton, NJ 08544, USA (email:
vaggarwa@princeton.edu). A. Sabharwal is with the Department of Electrical and Computer Engineering, Rice University, Houston, TX, 77005, USA (email:
ashu@rice.edu).}
 }

\date{}
\begin{document}
\maketitle
\begin{abstract}
Most communication systems use some form of feedback, often related to channel state information. The common models used in analyses either assume perfect channel state information at the receiver and/or noiseless state feedback links.
 However, in practical systems, neither is the channel estimate known perfectly at the receiver and nor is the feedback link perfect. In this paper, we study the  achievable diversity multiplexing tradeoff  using i.i.d. Gaussian codebooks, considering the errors in training the receiver and the errors in the feedback link for FDD systems, where the forward and the feedback are independent MIMO channels.

Our key result is that the maximum diversity order with one-bit of feedback information is identical to systems with more feedback bits. Thus, asymptotically in $\mathsf{SNR}$, more than one bit of feedback does not improve the system performance at constant rates. Furthermore, the one-bit diversity-multiplexing performance is identical to the system which has perfect channel state information at the receiver along with noiseless feedback link. This achievability uses novel concepts of power controlled feedback and training, which naturally surface when we consider imperfect channel estimation and noisy feedback links.
In the process of evaluating the proposed training and feedback protocols, we find an asymptotic expression for the joint probability of the $\mathsf{SNR}$ exponents of eigenvalues of the actual channel and the estimated channel which may be of independent interest.
\end{abstract}

\section{Introduction}

Channel state information at the transmitter has been well established to improve communication performance, measured either as increased capacity~(see e.g. \cite{GoldsmithVaraiya,CaireShamai,kuh08,book06,manish,cairepaper}), improved diversity-multiplexing performance~\cite{fsicc05,Farbod_Dissertation,steger,kim07,gamal06,kim08,other,newIT,steger,gaj,kim07,vaneet,vaneetno,gamal06,jaf08} or higher signal-to-noise ratio at the receiver~\cite{narula98,mukka03,book06} among many possible metrics. Several of the aforementioned works have considered the impact of incomplete channel knowledge at the transmitter, by considering quantized channel information at the transmitter which can be visualized to be made available by the receiver through a noiseless finite-capacity feedback link. While the model and subsequent analysis clearly shows that reduced channel information at the transmitter can lead to significant performance gains due to channel knowledge, a key requirement is that the receiver knows what the transmitter knows (even if there is an error in feedback link). In practice, the communicating nodes are distributed and have no way of aligning their channel knowledge perfectly.

In this paper, we systematically analyze the impact of mismatch in channel knowledge at the transmitter and receiver. For clarity of presentation, we will largely focus on a single-user point-to-point link (one transmitter and one receiver) and only at the end of the paper,  extend the results to the case of multiple-access channels (many transmitters and one receiver). The key departure from prior work is that we explicitly model both the forward and feedback links as fading wireless channels. A little thought immediately shows that all practical wireless networks have ``two-way" communication links, that is, the nodes are transceivers such that all the received and transmitted packets (control, data, feedback) travel over noisy fading channels. The two-way model to analyze feedback was first proposed in~\cite{steger} for TDD systems, where it was shown to enable accurate resource accounting of the feedback link resources (power and spectrum) \emph{and} analyze the important case of transmitter and receiver channel knowledge mismatch. A key message from~\cite{steger} was that in the general case when the transmitter and receiver knowledge is mismatched, both feedback and forward communication has to be \emph{jointly} designed. The simple two-way training protocol proposed in~\cite{steger}  led to the concept of power-controlled training which enabled joint estimation of transmitter's intended power control and actual channel realization.  We will continue the line of thought initiated in~\cite{steger} for the case of FDD (Frequency Division Duplex) systems and more importantly, systematically show how mismatch between information at the transmitter and receiver impacts the overall system performance.

For the case of FDD systems, we will continue to model the forward and feedback links as fading channels. However, unlike~\cite{steger}, the forward and feedback channel will be assumed to be completely uncorrelated since they use different frequency bands. In this case, the two-way training protocol proposed in~\cite{steger} will not be applicable and we will use the quantized feedback model, in which the receiver sends a quantized version of its own channel information over the feedback link.

Our  main result is an achievable diversity-multiplexing tradeoff for an $m\times n$ MIMO (Multiple Input Multiple Output) system, with measured channel information at the receiver which is used to send quantized channel information via a noisy link  using i.i.d. Gaussian codebooks. We show that, in the scope of modeling assumptions, diversity-multiplexing tradeoff with one-bit of noisy feedback with estimated CSIR (channel state information at receiver)  is identical to diversity-multiplexing tradeoff with perfect CSIR and noiseless one-bit feedback. Even more importantly, more bits of feedback do \emph{not} improve the maximum diversity order in the imperfect system (estimated CSIR with noisy feedback) in contrast of the idealized system where the maximum diversity order increases exponentially~\cite{fsicc05,Farbod_Dissertation,kim07}. The conclusion holds for multiuser systems in general.

The encouraging news from our analysis is two-fold. First, even noisy and mismatched information about the channel at both transmitter and receiver is sufficient to improve the whole diversity-multiplexing tradeoff when compared to the case with no feedback. Second, very few bits of feedback, in fact one-bit, is all one needs to build in a practical system.
However, to achieve any diversity order gain over the no-feedback system, a straightforward application of the Genie-aided feedback analysis~\cite{fsicc05,Farbod_Dissertation,kim07} fails. That is, conventional training followed by conventionally coded quantized feedback signal leads to no improvement in diversity order. A combination of power-controlled training (much like in~\cite{steger}) and the new concept of power-controlled feedback (proposed in this paper) appear essential for significant improvement in diversity-multiplexing gain compared to the no-feedback system.

We will build our main result in three major steps. In the first step, we assume that the receiver knowledge of the channel is perfect and the feedback channel is error-prone. Thus, the transmitter information about the channel is potentially different from that sent by the receiver. In this case, the coding of the feedback signal becomes important. If the feedback information about the channel is sent using a codebook where each codeword has equal power, then the maximum diversity order is $2mn$ irrespective of the amount of feedback information. In contrast, $b$-bit noiseless feedback leads to a maximum diversity order grows exponentially in number of bits~\cite{kim07,Farbod_Dissertation}. The reason for such dramatic decrease (from unbounded growth with $b$ in the noiseless case to maximum $2mn$ in the noisy case) is that the transmitter cannot distinguish between rare channel states, especially those where it is supposed to send large power to overcome poor channel conditions. As a result, it becomes conservative in its power allocation and the gain from power control does not increase unboundedly as in the case of noiseless feedback. A careful examination of the outage events on the \emph{feedback path} naturally points to unequal error protection in the form of  \emph{power controlled feedback} coding scheme, where the rare states are encoded with higher power codewords. With power controlled feedback, the diversity order can be improved to $(mn)^2 + 2mn$, a substantial increase compared to constant power feedback.

As our second step, we assume that the channel knowledge at the receiver is imperfect but the feedback path is error-free. We first analyze the commonly used protocol setup, where the receiver is trained at the average available power $\mathsf{SNR}$ (assuming noise variance is one) to obtain a channel estimate and then a quantized version of the estimated channel is fed back to the transmitter. While both the transmitter and receiver have identical information, the error in receiver information leads to \emph{no} gain in the diversity order compared to a no-feedback system. We take a cue from an earlier work in~\cite{steger} and the power-controlled feedback mechanism, and propose a two-round training protocol, where the receiver is trained twice. First round training uses an average power $\mathsf{SNR}$ and then after obtaining the channel feedback, the receiver is trained again with a training power dependent on the channel estimate. This implies that in poor estimated channel conditions, second round training power is higher than the first and in good conditions, it is lower. The adaptation of training power is labeled \emph{power-controlled training} and allows a substantial increase in maximum diversity order ($(mn)^2+mn$) compared to no-feedback case ($mn$). For general multiplexing gains, the achievable diversity order is identical to that obtained when the receiver knows perfect channel state information with 1 bit of perfect feedback in \cite{kim07}. Further, additional bits of feedback do not help in improving the maximum diversity order at constant rates when the receiver is trained to obtain a channel estimate.

Finally, we put the first two steps together to analyze the general case of noisy channel estimates with noisy feedback. The key conclusion is that the receiver channel estimate errors are the bottleneck in the maximum achievable diversity order. Thus, the maximum achievable diversity order at multiplexing gain $r=0$ is $(mn)^2+mn$, which is less than $(mn)^2+2mn$ obtained in the case of perfect receiver information. For general multiplexing, we get the same tradeoff as with $1$ bit of perfect feedback and the receiver having perfect channel state information as in \cite{kim07}. This diversity multiplexing tradeoff can be achieved with a single bit of noisy feedback from the receiver.

The results on MIMO point-to-point channels are extended to the multiple access channel in which all the transmitters have $m$ antennas each and the receiver has $n$ antennas, and all the conclusions drawn earlier for the single user also holds in a multiple access system. The channel information for each transmitter is measured at the receiver and a global quantized feedback is sent from the receiver to all the transmitters. We use the similar combination of power control training and feedback as in single-user systems to achieve a diversity order of $(mn)^2+mn$ with a single bit of power-controlled feedback in the main case in which the errors in the channel estimate and the feedback link are accounted.

We note that the improvement in diversity order is completely dependent on our use of power control as adaptation mechanism. For example, in MIMO systems, feedback can be used for beamforming~(e.g. \cite{narula98,mukka03}). However, beamforming does not lead to any change the diversity order and only increases the receiver $\mathsf{SNR}$ by a constant amount. Since we have focused on the asymptotic regime, we do not consider schemes like beamforming which have identical diversity-multiplexing performance as a non-feedback based system. In this paper, we find new achievable schemes to improve the diversity multiplexing tradeoff which are better than the traditional approaches but do not claim globally optimality of these schemes.

The rest of the paper is organized as follows. Section II describes preliminaries on channel model and diversity multiplexing tradeoff. Section III summarizes the known results for the case when the receiver knows perfect channel state information and the feedback is sent over a noiseless channel \cite{kim07}. Section IV describes the diversity order when the receiver knows the channel perfectly while the transmitter receives feedback on a noisy channel. Section V describes the diversity multiplexing tradeoff when the receiver is trained to get channel estimate while the feedback link to the transmitter is noiseless. Section VI consider both the above errors, i.e., it considers receiver estimating the channel and the feedback link is also noisy. Section VII presents some numerical results. We consider the extension to multiple access channels in Section VIII. Section IX concludes the paper.



\section{Preliminaries}

\subsection{Two-way Channel Model \label{sec:channel model}}

We will primarily focus on single-user multiple input-output channel with the transmitting node denoted by {\sf T} and receiving node denoted by {\sf R}. Later, we will extend the main result to the multiple access channel in Section~\ref{sec:multiple_access}. For the single-user channel, we will assume that there are $m$ transmit antennas at the source node and $n$ receive antennas at the destination node, such that the input-output relation is given by
\begin{equation}
\mathsf{T} \rightarrow \mathsf{R}: Y =  HX+ W,
\end{equation}
where the elements of  $H$ and $W$ are assumed to be i.i.d.\ with complex Normal distribution of zero mean and unit variance, $CN(0,1)$. The matrices $Y, H, X$ and $W$ are of dimension $n \times T_{\rm coh}, n\times m, m \times T_{\rm coh}$ and $n \times T_{\rm coh}$, respectively. Here $T_{\rm coh}$ is coherence interval such that
the channel $H$ is fixed during a fading block of $T_{\rm coh}$ consecutive channel uses, and statistically independent from one block to another.
The transmitter is assumed to be power-limited, such that the long-term power
is upper bounded, i.e, $\frac{1}{T_{\rm coh}}{\rm trace}({\mathbb E}\left[ XX^\dagger \right]) \leq
\mathsf{SNR}$.

Since our focus will be studying feedback over noisy channels, we assume that the same multiple antennas at the transmitter and receiver are available to send feedback on an orthogonal frequency band. For the feedback path, the receiver will act as a transmitter and the transmitter as a receiver. As a result, the feedback source (which is destination for data bits) will have $n$ transmit antennas and feedback destination (which is source of data bits) will be assumed to have $m$ receive antennas. Furthermore, a block fading channel model is assumed
\begin{equation}
\mathsf{R} \rightarrow \mathsf{T}:Y_f = H_f X_f + W_f,
\end{equation}
where $H_f$ is the MIMO fading channel for the feedback link, and the $W_f$ is the additive noise at the receiver of the feedback; both are assumed to have i.i.d.\ $CN(0,1)$ elements. The feedback transmissions are also assumed to be power-limited with a long-term power constraint given by $\frac{1}{T_{\rm coh}}{\rm trace}({\mathbb E}\left[ X_f X_f^\dagger \right]) \leq
\mathsf{SNR}_f$. Without loss of generality, we will assume the case where the transmitter and receiver have  \emph{symmetric resources}, such that $\mathsf{SNR} = \mathsf{SNR}_f$.

We note that a phase-symmetric two-way channel model with $H = H^T_f$ was studied earlier in~\cite{steger,gaj}. The phase-symmetric two-way channel is a good model for slow-fading time division duplex (TDD) systems. On the other hand, the above SNR-symmetric ($\mathsf{SNR} = \mathsf{SNR}_f$) model is well-suited for symmetrically resourced FDD systems. 

\begin{figure}[htbp]
\centering \myincludegraphics[height=3cm]
{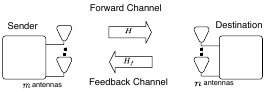}\caption{Two-way MIMO channel. The forward and feedback channels use the same antennas. If the forward and feedback are frequency-division duplex (FDD), then $H$ and $H_f$ are not equal. On the other hand, if the forward and feedback channels are time-division duplex (TDD), then $H = H^T_f$ is often a reasonable assumption.}\label{fig:channelmodel}
\end{figure}

%

\subsection{Obtaining Channel State Information}

\begin{figure}[htbp]
\centering \myincludegraphics[height=2.4cm]
{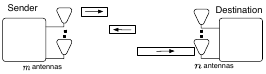}\caption{Two-way protocols: In a two-way channel, the protocols can involve many exchanges between the transmitter and receiver. The arrows indicate the direction of transmission.}\label{fig:two-way-protocol}
\end{figure}

The two-way channel model allows two-way protocols, as depicted in Figure~\ref{fig:two-way-protocol}, where the transmitter and receiver can conduct multiple back-and-forth transactions to complete transmission of one codeword. The model was used in~\cite{steger,gaj} to study the diversity-multiplexing performance of two-way training method. In this paper, we will focus on another instance of the two-way protocols which will involve channel estimation and quantized feedback. However, unlike~\cite{steger,gaj}, we will \emph{not} account for resources spent on feedback path since the methods developed in~\cite{steger} directly apply to the current case. Instead, we will focus on the more important issue of understanding how the mismatch in the transmitter and receiver channel information affects system performance.

To develop a systematic understanding of the two-way fading channel described in Section~\ref{sec:channel model}, we will consider two forms of receiver knowledge about the channel $H$. The receiver will either be assumed to know the channel $H$ perfectly or have a noisy estimate $\widehat{H}$ obtained using a minimum mean-squared channel estimate~(MMSE) via a training sequence (described in detail in Section~\ref{imcs}).

For the transmitter knowledge, the receiver will quantize its own knowledge ($H$ or $\widehat{H}$) and map it to an index $J_{\mathsf{R}}$, where $J_{\mathsf{R}} \in \{ 0, 1, \ldots, K-1\}$ where $K\ge 1$ is the number of feedback levels used by the receiver for feedback. The receiver then transmits the quantized channel information $J_{\mathsf{R}}$ over the feedback channel and the transmitter knowledge of the index is denoted by $J_{\mathsf{T}}$. For the case when the feedback channel is assumed to be noiseless, $J_{\mathsf{T}} = J_{\mathsf{R}}$, else $J_{\mathsf{T}} \neq J_{\mathsf{R}}$ with finite probability. Here the error probability will be depend on the channel signal-to-noise ratio $\mathsf{SNR}$ and the transmission scheme. Based on the quantized information about the channel, in the form of $J_{\mathsf{T}}$, the transmitter adapts its codeword to minimize the probability of outage (defined in the next section).

\subsection{Diversity-Multiplexing Tradeoff}

We will consider the case when then the codeword $X$ spans a single fading block. Based on the transmitter channel knowledge $J_{\mathsf{T}} = i$, the transmitted codeword is chosen from the codebook $C_i = \{X_{i}(1), X_{i}(2), \cdots , X_{i}(2^{RT_{\rm coh}})\}$, where the codebook rate is $R$. All $X_{i}(k)$'s (for $1\le k \le 2^{RT_{\rm coh}}$) are
matrices of size $m \times T_{\rm coh}$. We assume that $T_{\rm coh}$ is finite and does not scale with $\mathsf{SNR}$. In this paper, we will only consider single rate transmission where
the rate of the codebooks does not depend on the feedback index. Furthermore, we will assume codebooks $C_i$ are derived from the same base codebook $C$ by power scaling of the codewords. In other words $C_i = \sqrt{P_i} C$, where the product implies that each element of every codeword is multiplied by $\sqrt{P_i}$ where each element of  codeword $X(k) \in C$ has unit power. Thus,  $P_i$ is the power of the transmitted codewords. Recall that there is an average transmit power constraint, such that ${\mathbb E}(P_i) \leq \mathsf{SNR}$. We further assume that the base codebook $C$ consists of Gaussian entries, in other words we only focus on Gaussian inputs in this paper.

In point-to-point channels, outage is defined as the event that the mutual information of the channel for a channel knowledge $H\in\xi$ with some distribution $P_{H|\xi}$ (for $\xi$ a subset of all $n\times m$ matrices) at the receiver, $I_{\xi}(X;Y)$ is less than the desired rate $R$ \cite{zhengnon}. If $\xi$ is a singleton set containing $H$, the channel $H$ is known at the receiver in which case $I_\xi(X;Y) = \log\det\left(I+\frac{P(\jt)}{m}HQH^\dagger\right)$ is the mutual information of a point-to-point link with $m$ transmit and $n$ receive
antennas, transmit signal to noise ratio $P(\jt)$ and input distribution Gaussian with covariance matrix $Q$~\cite{zheng03}. The dependence of the index at the transmitter is made explicit by writing the transmit $\mathsf{SNR}$ as a function of $\jt$. Note that $P_i$ and $P(i)$ will mean the same thing in this paper.




Let $\Pi({\cal O})$ denote the probability of outage, where ${\cal O}$ is the set of all the channels where the maximum supportable rate $I_\xi(X;Y)$ is less than the transmitted rate $R$~\cite{zhengnon}.
The system is said to have diversity order of $d$ if
$\Pi({\cal O}) \doteq \mathsf{SNR}^{-d}$ \footnote{We adopt the notation
of \cite{zheng03} to denote $ \buildrel.\over=$ to represent
exponential equality. We similarly use $ \buildrel.\over<$, $
\buildrel.\over>$, $ \buildrel.\over\le$, $ \buildrel.\over\ge$ to
denote exponential inequalities.}. Note that all the index mappings, codebooks, rates, powers are
dependent on the average signal to noise ratio, $\mathsf{SNR}$. Specifically, the dependence of rate $R$ on
$\mathsf{SNR}$ is explicitly given by $R = r \log \mathsf{SNR}$, where $r$ is labeled as the multiplexing gain.
The diversity-multiplexing tradeoff is then described as the maximum diversity order $d(r)$ that can be achieved for a given multiplexing gain $r$.

The following result captures the diversity-multiplexing tradeoff for the case of perfect receiver information and no transmitter information. For a given rate $R \doteq r\log
\mathsf{SNR}$ and power $P\buildrel.\over= \mathsf{SNR}^{p}$, define the outage set ${\cal O}(R,P) = \{ H : \left(I+\frac{P}{m}HQH^\dagger\right)<R \}$. Denote the result diversity order as $G(r,p)$, which is $\Pi({\cal O}(R,P)) \buildrel.\over= \mathsf{SNR}^{-G({
r},{p})}$. The following result completely characterizes $G(r,p)$ and is a straightforward extension of the result in~\cite{zheng03}.




\begin{lemma}[\cite{vaneetno}]
\label{outlemma}
The diversity-multiplexing tradeoff for $P \buildrel.\over= \mathsf{SNR}^p$ and $r\le p\min(m,n)$ is given by the $(r,G(r,p))$ where
\[
G(r,p) =
 \inf
\mathop \sum
\limits_{i=1}^{\min(m,n)}(2i-1+\max(m,n)-\min(m,n))\alpha_i
\]
where the $\inf$ is over all $\alpha_1, \cdots, \alpha_{\min(m,n)}$ satisfying
\[
\{\alpha_1 \ge \ldots \alpha_{\min(m,n)}
\ge 0, \mathop \sum \limits_{i=0}^{\min(m,n)}(p-\alpha_i)^+<r\}.
\]
\end{lemma}
\begin{remark} {\rm
$G(r,p)$ is a piecewise linear curve connecting the points $(r,G(r,p))$=
$(kp,p(m-k)(n-k))$, $k = 0, 1, \ldots, \min(m,n)$ for fixed $m$, $n$ and $p>0$. This follows
directly from Lemma 2 of \cite{kim07}. $G(r,1)$ is the diversity multiplexing tradeoff with perfect CSIR and the transmit signal to noise ratio $\doteq\mathsf{SNR}$ \cite{zheng03}. Further, $G(r,p)=pG(\frac{r}{p},1)$.}
\end{remark}
%

\subsection{Summary of Results}

We will study the following four systems with different  accuracy of channel state information~(CSI) at the transmitter and receiver.

\begin{enumerate}
\item CSIR$\text{T}_{\text{q}}$: In this case, the receiver knowledge about $H$ is assumed to be perfect and the transmitter is assumed to receive a noiseless quantized feedback from the receiver about channel $H$. This case was first studied in~\cite{sk} and the diversity order increase was first proved in~\cite{fsicc05} with later extensions in~\cite{kim07}. The channel quantizer maps $H$ to an index $J_{\mathsf{R}}$ which is then communicated over the noiseless feedback channel. Since the feedback is noiseless, $J_{\mathsf{T}} = J_{\mathsf{R}}$. Based on the index $J_{\mathsf{T}}$, the transmitter adapts its transmission power  as described in Section~\ref{sec:all perfect}. The diversity order for $K$ levels of feedback is defined recursively as $d_{\text{R}\text{T}_\text{q}}(r,K) = G(r,1+d_{\text{R}\text{T}_\text{q}}(r,K-1))$ where $d_{\text{R}\text{T}_\text{q}}(r,0)=0$ and grows exponentially in number of feedback bits.

\item  CSIR$\widehat{\text{T}}_{\text{q}}$: Our first set of results analyze the case of imperfect channel knowledge at the transmitter, where the errors are caused by errors in the received quantized feedback index. In this case, $J_{\mathsf{T}} \neq J_{\mathsf{R}}$ with finite probability. We show that if a MIMO scheme optimized for equally likely input messages is used to send the feedback information, then the diversity order is limited to $d_{\text{R}\widehat{\text{T}}_\text{q}}(r,K)$  given in Table~\ref{tbl:all div-mux tradeoffs}.
However, the feedback information is not equally-likely. Hence, we propose a natural unequal error protection method labeled \emph{power controlled feedback}, where the power control is performed based on input probabilities. The power-controlled feedback results in a diversity-order $\overline{d}_{\text{R}\widehat{\text{T}}_\text{q}}(r,K)$, specified in Table~\ref{tbl:all div-mux tradeoffs}.
With power-controlled feedback, the maximum diversity order increases from $d_{\text{R}\widehat{\text{T}}_\text{q}}(0,K) = 2mn$ to $\overline{d}_{\text{R}\widehat{\text{T}}_\text{q}}(0,K) = mn(mn+2)$ for all $K > 2$. For $K=2$, the maximum diversity order increases from $d_{\text{R}\widehat{\text{T}}_\text{q}}(0,K) = 2mn$ to $\overline{d}_{\text{R}\widehat{\text{T}}_\text{q}}(0,K) = mn(mn+1)$.

\item CSI$\widehat{\text{R}}\text{T}_{\text{q}}$: Our next step will be to isolate the effect of errors in receiver knowledge of the channel. Thus, the receiver will estimate $\widehat{H}$, which will be mapped to
${J}_{\mathsf{R}}$. Since the feedback is assumed to be perfect, the transmitter knowledge is same as that of receiver, $J_{\mathsf{T}} = {J}_{\mathsf{R}}$.
We will show that for a three-phase power-controlled training based protocol can achieve a diversity order of $\overline{d}_{\widehat{\text{R}}\text{T}_q}(r,K) =
G(r,1+G(r,1))$. In fact, if the power-controlled training is not performed then channel estimation errors completely dominate and feedback is rendered useless; the resultant tradeoff collapses to $d_{\widehat{\text{R}}\text{T}_q}(r,K) = G(r,1)$ for all $K$.
So analogous to power-controlled feedback, power-controlled training appears essential to improve the diversity order in this case.

\item CSI$\widehat{\text{R}}\widehat{\text{T}}_{\text{q}}$: Finally, we put the above two cases together and derive the maximum achievable diversity order as $\overline{d}_{\widehat{\text{R}}\widehat{\text{T}}_q}(r,K)  = G(r,1+G(r,1))$, which is equivalent to the case with only receiver errors.
Thus, we conclude that the receiver errors dominate the achievable diversity order.

\end{enumerate}
\begin{remark}\label{rem1}
An important word of caution for all the results (new and known) summarized in Table~\ref{tbl:all div-mux tradeoffs}. Unlike the previous work in~\cite{zhengnon,steger}, we do \emph{not} account for the resources spent in channel training and feedback in this paper. Resource accounting can be performed using the procedure developed in~\cite{steger}, by scaling the multiplexing $r$ appropriately. More precisely, the multiplexing $r$ should be replaced by $rT_{\rm coh}/(T_{\rm coh}-1)$ for CSIR$\widehat{\text{T}}_{\text{q}}$, $rT_{\rm coh}/(T_{\rm coh}-2m)$ for the three-phase protocol in power-controlled training of CSI$\widehat{\text{R}}\text{T}_{\text{q}}$, $rT_{\rm coh}/(T_{\rm coh}-m)$ for  two-phase protocol in constant power training of CSI$\widehat{\text{R}}\text{T}_{\text{q}}$, and $rT_{\rm coh}/(T_{\rm coh}-2m-1)$ for the three-phase protocol of CSI$\widehat{\text{R}}\widehat{\text{T}}_{\text{q}}$. The resource accounting multipliers assumes that the feedback requires one channel use and the training requires $m$ channel uses. The details are further explained in Remarks~3 and~5. Further, the use of the number of antennas to use and the protocols would need to be optimized for each multiplexing as in \cite{zhengnon,steger} and is omitted in this paper for readability.
\end{remark}
\begin{table*}
\begin{center}
\caption{Summary of Diversity-Multiplexing Tradeoffs. See Remark \ref{rem1} for a caution in using this table.
\label{tbl:all div-mux tradeoffs}}
\begin{tabular}{|c|c|c|} \hline
Case & Main Characteristic & D-M Tradeoff \\ \hline
CSIRT & Perfect Information at ${\sf T}$ and {\sf R} & $d_{\text{RT}}(r) = \infty \:\: \forall r < \min(m,n)$ \\ \hline
CSIR$\text{T}_{\text{q}}$ & Quantized Information at ${\sf T}$ & $d_{\text{R}\text{T}_\text{q}}(r,K) = G(r,1+d_{\text{R}\text{T}_\text{q}}(r,K-1))$,  $d_{\text{R}\text{T}_\text{q}}(r,0)=0$ \\ \hline \hline
CSIR$\widehat{\text{T}}_{\text{q}}$ & Noisy information at {\sf T} & $d_{\text{R}\widehat{\text{T}}_\text{q}}(r,K) = \min(\overline{B}_{K}({ r}),
mn+G(r,1))$ \\
& (Constant Power Feedback) & \\ \hline
CSIR$\widehat{\text{T}}_{\text{q}}$ & Noisy information at {\sf T} & $\overline{d}_{\text{R}\widehat{\text{T}}_\text{q}}(r,K) = \min\left(d_{\text{R}\text{T}_\text{q}}(r,K),\max_{q_j\le1+d_{\text{R}\text{T}_\text{q}}(r,j)}
\right.$\\& (Power-controlled Feedback) &$\left.\min_{i=1}^{K-1}(mn((q_i)^+-(q_{i-1})^+)+d_{\text{R}\text{T}_\text{q}}(r,i)) \right)$  \\ \hline \hline
CSI$\widehat{\text{R}}\text{T}_{\text{q}}$ & Noisy information at {\sf R} &   $d_{\widehat{\text{R}}\text{T}_q}(r,K) = G(r,1)$  \\
& (Constant Power Training) &   \\ \hline
 CSI$\widehat{\text{R}}\text{T}_{\text{q}}$& Noisy information at {\sf R} & $\overline{d}_{\widehat{\text{R}}\text{T}_q}(r,K)=G(r,1+G(r,1))$  \\
& (Power-controlled training) & \\  \hline \hline
CSI$\widehat{\text{R}}\widehat{\text{T}}_{\text{q}}$ & No Genie-aided information & $d_{\widehat{\text{R}}\widehat{\text{T}}_q}(r,K) = G(r,1+G(r,1))$  \\
& (Power-controlled training \& feedback) & \\ \hline
\end{tabular}
\end{center}
\end{table*}


\section{CSIR$\text{T}_{\text{q}}$: Perfect CSIR with Noiseless Quantized Feedback \label{sec:all perfect}}

The diversity-multiplexing tradeoff for the case of receiver with perfect information and noiseless quantized information has been extensively studied  in \cite{Farbod_Dissertation,kim07,vaneet}. In this section, we will discuss the main ideas for the single user MIMO channel model stated in Section~\ref{sec:channel model}. We start off with an example to illustrate the main idea.\\

\begin{figure*}[htbp]
\centering \myincludegraphics[height=5cm]
{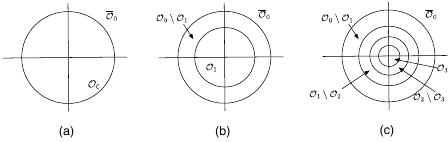}\caption{Example 1: Channel events for the case of (a) no feedback (${\cal O}_0$ is the set of channels in outage) and (b) one-bit feedback (${\cal O}_1$ is the outage set) and (c) two-bit feedback (${\cal O}_3$ is the outage set).}\label{fig:one-bit-feedback}
\end{figure*}

\noindent
\emph{Example 1 (SISO)}: Consider the case where $m=n=1$. Without any feedback, the maximum diversity (at $r \to 0$) is 1. The space of channels $\{H\} = {\mathbb C}$ can be divided into two sets: the outage set ${\mathcal O}_0 = \left\{ H : \|H\|^2 < \frac{2^R-1}{\mathsf{SNR}} \right\}$ and its complement $\overline{\mathcal O}_0 = \{ H \} \setminus {\mathcal O}_0$; see Figure~\ref{fig:one-bit-feedback}(a).
The probability of the set $\Pi({\mathcal O}_0) \approx \mathsf{SNR}^{-1}$ and $\Pi(\overline{\mathcal O}_0) \approx (1-\mathsf{SNR}^{-1})$.

With one-bit of noiseless feedback, the receiver can convey whether the current channel $H$ belongs to ${\mathcal O}_0$ or $\overline{\mathcal O}_0$. Since the outage event is rare, the transmitter can send much larger power than usual to reduce the outage probability as follows. When the feedback index is $\jr = 1$ representing event ${\mathcal O}_0$, the transmitter will use transmit power $\frac{1}{2}\mathsf{SNR}/\Pi({\mathcal O}_0)\approx \frac{1}{2}\mathsf{SNR}^{2}$. For the feedback index $\jr = 0$, representing $\overline{\mathcal O}_0$, power $\frac{1}{2}\mathsf{SNR}$ is used. The average power used in the above two-level power control is $\frac{1}{2} \mathsf{SNR}/\Pi({\mathcal O}_0) \cdot \Pi({\mathcal O}_0) + \frac{1}{2} \mathsf{SNR} \cdot (1-\mathsf{SNR}^{-1}) \le \mathsf{SNR}$. With the above one-bit power control, the outage probability is $\approx \mathsf{SNR}^{-2}$ since the set of channels in outage is reduced to ${\mathcal O}_1 = \left\{ H : \|H\|^2 < \frac{(2^R-1)\cdot 2\cdot \Pi({\mathcal O}_0)}{\mathsf{SNR}} \right\} \approx \left\{ H : \|H\|^2 < \frac{(2^R-1)\cdot 2}{\mathsf{SNR}^2} \right\}$; see Figure~\ref{fig:one-bit-feedback}(b).

If the feedback rate was $\log_2(3) \text{ bits/channel state}$, then the receiver could convey information about three events $\{ {\mathcal O}_1, {\mathcal O}_0 \setminus {\mathcal O}_1, \overline{\mathcal O}_0 \}$. In state ${\mathcal O}_1$, the transmitter can send power $\approx \mathsf{SNR}^3$ since $\Pi({\cal O}_1) \approx \mathsf{SNR}^{-2}$ and thus reduce the outage probability to $\mathsf{SNR}^{-3}$. If we had two bits of feedback or equivalently $K=4$ levels, then the diversity order will be 4 and the outage region will be given by ${\cal O}_3$ which will be the set of all those channels which could not support rate $R$ with power $\approx \mathsf{SNR}^{4}$; see Figure~\ref{fig:one-bit-feedback}(c). Using the above recursive argument, for $K$ levels of feedback, a diversity order of $K=2^{b}$ can be achieved, where $b$ is the number of feedback bits per channel realization. \hfill $\blacksquare$

The above example captures the essence of the general result for the MIMO channels, given by the following theorem.

\begin{theorem}[\cite{kim07}]Suppose that $K\ge1$ and $r<\min(m,n)$. Then, the diversity-multiplexing tradeoff for the case of perfect CSIR with noiseless quantized feedback is $d_{\text{R}\text{T}_\text{q}}(r,K)$  defined recursively as \[
d_{\text{R}\text{T}_\text{q}}(r,K) =
G(r,1+ d_{\text{R}\text{T}_\text{q}}(r,K-1)),\]
where $d_{\text{R}\text{T}_\text{q}}(r,0)=0$.
\label{th:perfect case}
\end{theorem}

The main idea of the proof is along the lines of Example~1 and is summarized as follows. Based on its knowledge of $H$, the receiver decides the feedback index $J_{\mathsf{R}} = J_{\mathsf{T}}  \in\{0,1,\cdots,K-1\}$, where $K=2^b$ is the number of quantization levels and $b$ is the number of feedback bits. If the transmitter receives feedback index $\jt$, it sends data at power level $P_{\jt}$. Without loss of generality, $P_0\le P_1\le\cdots\le P_{K-1}$. The optimal deterministic index mapping has the following form \cite{kim07},
\[
 \jr =
\begin{cases}
\arg \min_{i \in \mathbb{I}}  i,  & \mathbb{I} = \left\{ k : \log \det(I+{H}{H}^\dagger {P_{k}})\ge R,\right.\\ &\left. \quad  k \in \{0,\cdots,K-1\} \right\}  \\
  0,  & \text{ if the set $\mathbb{I}$ is empty } \\
\end{cases}.
\]

Based on the above assignment of feedback indices, the optimal power levels can be found out as $P_i\doteq\mathsf{SNR}^{1+p_i}$ where $p_i$ are recursively defined as: $p_0=0$, $p_j=G(r,1+p_{j-1})\forall j\ge 1$. Using the above recursion, the optimal diversity is given by $G(r,1+p_{K-1})$ which reduces to $d_{\text{R}\text{T}_\text{q}}(r,K)$, thus proving Theorem~\ref{th:perfect case}. We note that special cases of Theorem~\ref{th:perfect case} were also proved in~\cite{Farbod_Dissertation}.

We note that event that $\{H : \log \det(I+{H}{H}^\dagger {P_{K-1}})<R \}$ corresponds to the outage event, because none of the power levels can be used to reach a mutual information of $R$. Thus any state can be assigned to the outage event and in fact, the diversity-multiplexing tradeoff is unaffected by which index is used to represent this state. Since the receiver knowledge is perfect, using the index $J_{\mathsf{R}} = 0$ ensures that the overall power consumption is minimized. However, when the receiver or transmitter knowledge is not perfect (which will be the case in the rest of the paper), we will assign the perceived outage state the highest power level $P_{K-1}$, which helps reduce the outage probability due to misestimation of channel.

The maximum diversity order in Theorem~\ref{th:perfect case} increases very rapidly with the number of feedback levels $K$. The maximum diversity order~\cite{kim07} $d_{\text{R}\text{T}_\text{q}}(0,K) = \sum_{g=1}^K (mn)^g$, which grows exponentially fast in the number of levels $K$. In fact, as $K\rightarrow \infty$, the diversity order also increases unboundedly. Since as $K \rightarrow \infty$, the feedback approaches the perfect feedback and as a result, perfect channel inversion becomes possible. In~\cite{CaireShamai}, it was shown that for perfect channel state information at transmitter and receiver, zero outage can be obtained at finite SNR if $\max(m,n) > 1$. Perfect channel inversion essentially converts the fading channel into a vector Gaussian channel, whose error probability goes to zero for all rates less than its capacity. The unbounded growth of diversity order is rather unsettling and is in fact, a fragile result as shown by our results in the following sections.


\section{CSIR$\widehat{\text{T}}_{\text{q}}$: Perfect CSIR with Noisy Quantized Feedback\label{imfeed}}

In this section, we will analyze the case when the receiver knows the channel $H$ perfectly but the quantization index $J_{\mathsf{R}}$ is conveyed to the transmitter over a noisy feedback link, resulting in the event $J_{\mathsf{T}} \neq J_{\mathsf{R}}$ with non-zero probability. We will consider two feedback designs. In the first design, the feedback channel will use a constant power transmission scheme designed for equally-likely symbols (which is the commonly studied case for i.i.d.\ data). Learning from the limitations of constant power feedback transmission, we will then construct a new power-controlled feedback strategy which will exploit the unequal probability of different channel events quantized at the receiver.

\subsection{Constant Power Feedback Transmission \label{sec:constant power feedback}}

First, we observe the impact of using the power control described in Section~\ref{sec:all perfect} by considering Example~1.\\

\begin{figure*}[htbp]
\centering \myincludegraphics[width=6.5in]
{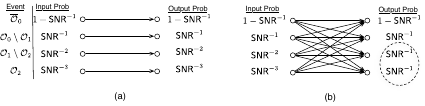}\caption{Example 2: Input-output probabilities of the feedback channel, when (a)~the channel is noiseless and (b)~when the channel is noisy causing mismatch between transmitter and receiver knowledge, assuming that a code optimized for equally-likely source is used. The noise in the feedback channel changes the output probabilities, as shown by the circled output probabilities. (Note that the sum of probabilities is more than one since we have omitted the constants for ease of understanding).}\label{fig:noisyfeed}
\end{figure*}

\noindent
\emph{Example 2 (Impact of Feedback Errors, SISO)}: Consider the case of $b=2$ feedback bits which allows $K=4$ feedback indices.
In this case, the receiver can convey four events, which we choose to be  ${\cal O} = \{ \overline{\cal O}_0, {\cal O}_0 \setminus {\cal O}_1 , {\cal O}_1 \setminus {\cal O}_2, {\cal O}_2 \}$~(shown in Figure~\ref{fig:one-bit-feedback}(c)), where the events are described as
\begin{equation}
{\cal O}_i = \left\{ H : \log\left(1+ |H|^2 \mathsf{SNR}^{1+i} \right) < R \right\}, i = 0,1,2.
\end{equation}
As discussed in the previous section, the power control in the forward channel uses four different power levels, $P_{\jt}\doteq \mathsf{SNR}^{1+\jt}$ for $\jt=0,1,2,3$.
The probability of each of the above events and the associated power control is defined in Table~\ref{tbl:probabilities}.
\begin{table*}
\begin{center}
\caption{Example 2: Probability of events at the transmitter and receiver with and without noise in the feedback channel. Receiver knowledge is assumed to be perfect. (Caution: Probabilities are only reported up to their order, and constants such that they sum to one are omitted.)\label{tbl:probabilities}}
\begin{tabular}{|c|c|c|c|c|} \hline
Event & Feedback & Prob of Event  & Prob at Transmitter  & Prob at Transmitter  \\
&Index ($J_{\mathsf{R}}$) & at Receiver & (noiseless feedback) & (noisy feedback) \\ \hline
$\overline{\cal O}_0$ & 0 & $(1-\mathsf{SNR}^{-1})$ &  $(1-\mathsf{SNR}^{-1})$ &  $(1-\mathsf{SNR}^{-1})$ \\
${\cal O}_0 \setminus {\cal O}_1$ & 1 & $\mathsf{SNR}^{-1}$ & $\mathsf{SNR}^{-1}$  &$\mathsf{SNR}^{-1}$ \\
${\cal O}_1 \setminus {\cal O}_2$ & 2 & $\mathsf{SNR}^{-2}$ & $\mathsf{SNR}^{-2}$  & $\mathsf{SNR}^{-1}$ \\
${\cal O}_2$ & 3 & $\mathsf{SNR}^{-3}$ & $\mathsf{SNR}^{-3}$ & $\mathsf{SNR}^{-1}$ \\ \hline
\end{tabular}
\end{center}
\end{table*}
The last column in Table~\ref{tbl:probabilities} shows the impact of probability of events as seen by the transmitter when there are errors in the feedback link such that $\Pi(\jt=j|\jr=i) \stackrel{\cdot}{=} \mathsf{SNR}^{-1}$ for $j\ne i$ as shown in Figure~\ref{fig:noisyfeed}(b). For example, with noisy feedback, $\Pi(J_{\mathsf T}=3)=\sum_{j=0}^3\Pi(J_{\mathsf T}=3|J_\mathsf{R}=j)\Pi(J_\mathsf{R}=j)\doteq \mathsf{SNR}^{-1}+\mathsf{SNR}^{-1}\mathsf{SNR}^{-1}+\mathsf{SNR}^{-1}\mathsf{SNR}^{-2}+\mathsf{SNR}^{-3}\doteq \mathsf{SNR}^{-1}$.

In fact, there are two dominant error events, first being $\overline{\cal O}_0$ ($\jr = 0$) being confused as either ${\cal O}_0 \setminus {\cal O}_1$ or ${\cal O}_1 \setminus {\cal O}_2$ or ${\cal O}_2$ ($\jt = 1,2$ or 3) which amounts to limiting the maximum power that can be used and the second being the probability that $\overline{\cal O}_1$ ($\jr = 1$) being confused as $\overline{\cal O}_0$ ($\jt = 0$).
The second error event has approximate probability of $\mathsf{SNR}^{-2}$ which limits the maximum diversity to $2$.
%
%
%
Thus, the event probabilities at transmitter are no longer the same in the presence of the feedback errors. As a result, the transmitter \emph{cannot} use the power control $P_{\jt} = \mathsf{SNR}^{1+\jt}$ without exceeding the average power constraint. In fact, the highest power the transmitter can send is of the order $\doteq\mathsf{SNR}^2$, and the maximum diversity order of the constant power feedback mechanism is limited to $2$ \emph{irrespective} of the number of feedback bits. \hfill $\blacksquare$

We now generalize the above example to MIMO for arbitrary multiplexing gain.
Define $\overline{B}_{j} ({ r} )$ be defined by the recursive equation
\begin{eqnarray}\label{ceq}
\overline{B}_{j} ({ r} ) =
\begin{cases}
 0, & j = 0 \\
 G({ r} ,1+\min(mn,\overline{B}_{j-1}({ r})) ), & j \ge 1 \\
 \end{cases} .\nonumber
\end{eqnarray}

\begin{theorem}\cite{vaneetno}
Suppose that $K>1$ and $r<\min(m,n)$. Then, the
diversity-multiplexing tradeoff for the constant power feedback transmission with $K$
indices of feedback is given by
\[
d_{\text{R}\widehat{\text{T}}_\text{q}}(r,K) =\min(\overline{B}_{K}({ r}),
mn+G(r,1)).\]
\end{theorem}
\begin{remark}
Accounting for the feedback resources can be done as follows. In the limit of high $\mathsf{SNR}$, the feedback will consume one channel use and hence to get the rate $R\doteq r\log(\snr)$, $r$ should be replaced by $rT_{\rm coh}/(T_{\rm coh}-1)$ in the above expression. However, the reader must note that there are implicit time factor terms which can be easily integrated and one may carry out an optimization on the diversity obtained as a function of these time loss terms for each multiplexing. For example, for $r>\min(m,n)(T_{\rm coh}-1)/T_{\rm coh}$, the feedback would not be useful due to the time spent and it is better to use a non-feedback based strategy.
\end{remark}
Much like in Example~2, the maximum diversity order for $r \rightarrow 0$ is $2mn$ for all $K \geq 2$. That is one-bit of feedback is sufficient to achieve the maximum diversity order with constant-power feedback. However, for $r > 0$, as the number of feedback levels $K$ increases, the diversity order $d_{\text{R}\widehat{\text{T}}_\text{q}}(r,K)$ also increases, such that  $d_{\text{R}\widehat{\text{T}}_\text{q}}(r,K) \leq G(r,1) + mn$ for all $r$. Note that $G(r,1)$ is the diversity order without any feedback. Coincidentally, the diversity order of the feedback link is $mn$ (feedback link is non-coherent where the error probability can decay no faster than $\mathsf{SNR}^{-mn}$), and ends up determining the maximum gain possible beyond $G(r,1)$.

The key bottleneck in the above result is that the feedback link is using a transmission scheme optimized for equally-likely signals, which is appropriate if the feedback link was being used to send equally-likely messages like data packets. However, the information being conveyed in the feedback link is not equiprobable and hence the usual MIMO schemes optimized for equally-likely messages are not well suited. Again, in the context of Example~2, we will first show an alternate design of \emph{power-controlled feedback} can improve the diversity order in the next section. Before we proceed, we note that our analysis in~\cite{steger} for a TDD two-way channel yields the same maximum diversity order of $2mn$. It is satisfying to see how two different feedback methods have identical behavior; their relationship is further explored in~\cite{vaneetasil}.

\subsection{Power-controlled Feedback \label{sec:power-controlled feedback}}

In this section, we exploit the unequal probabilities of the outage events at different power levels to develop a power-controlled feedback scheme to reduce overall outage probabilities. Our feedback transmission will be designed for a non-coherent channel, since the feedback channel $H_f$ is not known at the transmitter or the receiver.\\

\begin{figure}[htbp]
\centering \myincludegraphics[width=3.25in]
{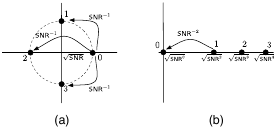}\caption{Examples 2 and 3: Dominant error events when (a)~the feedback channel uses a transmission suited for equally-likely messages and (b)~a power-controlled feedback design where codewords are power-controlled based on their probability of occurrence.}\label{fig:feedback-design}
\end{figure}

\noindent
\emph{Example 3 (Power-controlled Feedback)}: For the constant-power feedback transmission scheme described in the Section~\ref{sec:constant power feedback}, each input is mapped to a codeword with the same power and is pictorially depicted in Figure~\ref{fig:feedback-design}(a).
However, since the events in set ${\cal O}$ are not equi-likely, we assign different power levels to each event. Now consider the power assignments shown in Table~\ref{tbl:feedback}, also depicted in Figure~\ref{fig:feedback-design}(b) for the feedback channel. The average power over the feedback channel is
\begin{eqnarray}
&&\mathsf{SNR}^0 (1-\mathsf{SNR}^{-1}) + \mathsf{SNR}^2 \mathsf{SNR}^{-1} + \mathsf{SNR}^3 \mathsf{SNR}^{-2}\nonumber\\ &&+ \mathsf{SNR}^{-3} \mathsf{SNR}^{4} \doteq \mathsf{SNR}\nonumber
\end{eqnarray}
Here the rare events are conveyed with more power, which allows a more reliable delivery of the feedback information without violating feedback power constraint. Thus, the power-controlled scheme resembles an amplitude shift keying.
\begin{table*}
\begin{center}
\caption{Example 3: Power assignment for noisy feedback channel.  (Caution: Probabilities are only reported up to their order, and constants such that they sum to one are omitted.) \label{tbl:feedback}}
\begin{tabular}{|c|c|c|c|c|} \hline
Event & Feedback & Prob at Receiver &  Feedback Transmit & Prob at Transmitter  \\
& Index & & Power & (noisy feedback) \\ \hline
$\overline{\cal O}_0$ & 0 & $(1-\mathsf{SNR}^{-1})$ & $\mathsf{SNR}^{0}$ &  $(1-\mathsf{SNR}^{-1})$ \\
${\cal O}_0\setminus {\cal O}_1$ & 1 & $\mathsf{SNR}^{-1}$ &  $\mathsf{SNR}^{2}$ &$\mathsf{SNR}^{-1}$ \\
${\cal O}_1\setminus {\cal O}_2$ & 2 & $\mathsf{SNR}^{-2}$ &  $\mathsf{SNR}^{3}$ & $\mathsf{SNR}^{-2}$ \\
${\cal O}_2$ & 3 & $\mathsf{SNR}^{-3}$ &  $\mathsf{SNR}^{4}$ & $\mathsf{SNR}^{-3}$ \\ \hline
\end{tabular}
\end{center}
\end{table*}

The reason that the probabilities at the transmitter and the receiver are same for power-controlled is because the events $\jt>\jr$ occur with
exponentially small probability. The outage probability can be lower bounded by $ \Pi(\jt=0,\jr=1)$ which is an event when the receiver requested
a higher power level than what it received. Note that $\Pi(\jr=1)\doteq \mathsf{SNR}^{-1}$ and $\Pi(\jt=0|\jr=1)\doteq \mathsf{SNR}^{0-2}$.  Hence, $\Pi(J_{\mathsf{T}}=0,J_{\mathsf{R}}=1)\doteq \mathsf{SNR}^{-1}\mathsf{SNR}^{-2} = \mathsf{SNR}^{-3}$ is the probability of
this dominant error event. Note that both the forward and feedback channel power constraints are satisfied. Hence, we can obtain a diversity order of $3$,
which is higher than the diversity order of 2 obtained via constant-power feedback design (Example~2) but lower than~4 that can be achieved with a noiseless
 feedback channel (Example~1).\hfill $\blacksquare$

We will generalize the above example to the case of general MIMO channels by using the following power-control for the feedback channel. Let the feedback power level for different feedback messages be $Q_0\doteq\mathsf{SNR}^{q_0}$, $Q_1\doteq\mathsf{SNR}^{q_1}$, $\ldots$, $Q_{K-1}\doteq \mathsf{SNR}^{q_{K-1}}$ with $q_j>q_{j-1}$. Assume for now that the set of powers $\{ Q_i \}$ satisfy the feedback power constraint of $\mathsf{SNR}$.

We will use the maximum aposteriori probability~(MAP) detection rule to detect the message transmitted by the receiver on the feedback channel. If the transmitted signal is at signal to noise ratio level of $Q$, the received power is $S_f \doteq {\rm trace}(HH^\dagger Q+WW^\dagger+\sqrt{Q}(HW^\dagger+WH^\dagger))$. Since we are encoding the feedback index information in signal power, we will compare the received power $S_f$ with different threshold power levels. We first note that for all $q_i\le 0$, the received power will be dominated by the noise term ${\rm trace}(WW^\dagger)$, which is $\doteq 1$ with high probability.
Thus, we assume that $q_j>0$ for $j>0$.

Suppose the eigenvalues of $HH^\dagger$ are $(\lambda_1, \cdots, \lambda_{m_N})$ and $\lambda_i\doteq\mathsf{SNR}^{-\alpha_i}$ where $\alpha_i$ are the negative $\mathsf{SNR}$
exponents of the $\lambda_i$, $m_N=\min(m,n)$ and ${\ba}=(\alpha_1,\cdots,\alpha_{m_N})$. Then, the distribution of $\alpha_i$ is given by the following result.
\begin{lemma}\cite{zheng03}\label{zh} Assume $\alpha_1\ge \alpha_2\ge \cdots \alpha_{m_N}$ are the power exponents as described above. In the limit of high $\mathsf{SNR}$, the probability density function of the $\mathsf{SNR}$ exponents, $\ba$, of the eigenvalues of $HH^\dagger$ is given by
\begin{eqnarray}
p({\ba})&\doteq& \Pi_{i=1}^{m_N}\mathsf{SNR}^{-(2i-1+|n-m|)\alpha_i}{\mathbf 1}_{\min(\ba)\ge0}.
\end{eqnarray}
\end{lemma}

We now show that the thresholds for MAP decoding are $\mathsf{SNR}^{\max(q_0,0)+\epsilon}$, $\mathsf{SNR}^{q_1+\epsilon}$, $\ldots$,$\mathsf{SNR}^{q_{K-2}+\epsilon}$ for some small $\epsilon>0$.
This can be derived by observing reason is that $\Pi(\jt>i|\jr=i)$ decays faster than polynomially\footnote{Any function $f(\mathsf{SNR})$ which decays faster than any polynomial in $\mathsf{SNR}$, like exponential or super-exponential functions of $\mathsf{SNR}$, are $\doteq 0$ since $\lim_{\mathsf{SNR}\to \infty}\frac{log(f(x))}{\log(\mathsf{SNR})}=0$.} in $\mathsf{SNR}$ and is hence $\doteq 0$ as long as the threshold for detecting $\jt=i$ is above $\mathsf{SNR}^{q_i+\epsilon}$ for any $\epsilon>0$. To see this, for $0\le j<K-1$,
\begin{eqnarray}
&&\Pi(\jt>j|\jr=j)\nonumber\\
&=&\Pi\left({\rm trace}(HH^\dagger \mathsf{SNR}^{\max(q_j,0)})+1\ge\mathsf{SNR}^{\max(q_j,0)+\epsilon}\right)\nonumber\\
&\doteq& \Pi\left(\sum(\mathsf{SNR}^{q_j-\alpha_i})+1\ge\mathsf{SNR}^{\max(q_j,0)+\epsilon}\right)\nonumber\\
&\doteq& \Pi\left(\mathsf{SNR}^{\max(q_j-\min\alpha_i,0)}\ge\mathsf{SNR}^{\max(q_j,0)+\epsilon}\right)\nonumber\\
&\doteq& \Pi\left(\max(q_j-\min\alpha_i,0)\ge\max(q_j,0)+\epsilon\right)
\end{eqnarray}
We see that for the last expression to occur with finite probability, $\min\alpha_i<0$ which cannot happen with polynomially decreasing probability by Lemma \ref{zh} and hence $\Pi(\jt>j|\jr=j)\doteq 0$ irrespective of $\epsilon>0$. Now, for MAP detection, we would find a threshold between $q_i$ and $q_{i+1}$ so as to minimize  $\Pi(\jt=i+1,\jr=i)+\Pi(\jt=i,\jr=i+1)$. Since the first term $\doteq 0$ and the second term is minimized by choosing $\epsilon$ as small as possible, choosing $\epsilon$ small enough gives the desired threshold for MAP decoding.


Further, for all $Q\dot{>} \mathsf{SNR}$, ${\rm trace}(HH^\dagger Q+WW^\dagger+\sqrt{Q}(HW^\dagger+WH^\dagger))\doteq {\rm trace}(HH^\dagger)Q+1$. Hence,
 we obtain $\Pi(\jt=0|\jr=1)=\Pi({\rm trace}(HH^\dagger \mathsf{SNR}^{p_1})+1\dot{<}\mathsf{SNR}^{\max(p_0,0)+\epsilon})$. Let $\lambda_i$ be the eigenvalues of $HH^\dagger$ and $\lambda_i\doteq \mathsf{SNR}^{-\alpha_i}$. Then,
\begin{eqnarray}
&&\Pi(\jt=0|\jr=1)\nonumber\\&=&\Pi\left({\rm trace}(HH^\dagger \mathsf{SNR}^{q_1})+1\dot{<}\mathsf{SNR}^{\max(q_0,0)+\epsilon}\right)\nonumber\\
&\doteq& \Pi\left(\sum_{i=1}^{\min(m,n)}\mathsf{SNR}^{q_1-\alpha_i}+1<\mathsf{SNR}^{\max(q_0,0)+\epsilon}\right)\nonumber\\
&\doteq& \Pi\left(\mathsf{SNR}^{\max(q_1-\min\alpha_i,0)}\dot{<}\mathsf{SNR}^{\max(q_0,0)+\epsilon}\right)\nonumber\\
&\doteq& \Pi\left(\max(q_1-\min\alpha_i,0)<\max(q_0,0)+\epsilon\right)\nonumber\\
&\doteq& \Pi\left(\min\alpha_i>q_1-\max(q_0,0)-\epsilon\right)\nonumber\\
&\doteq& \mathsf{SNR}^{-mn(q_1-\max(q_0,0))}\nonumber
\end{eqnarray}
where the last step follows from Lemma \ref{zh}.
Since $q_1\le 1+G(r,1)$ (as  $\Pi(\jr=1)\dot{\ge}\mathsf{SNR}^{-G(r,1)}$ and there is power constraint for feedback link of $\mathsf{SNR}$), the outage probability is lower bounded by $ \Pi(\jt=0,\jr=1)\ge \mathsf{SNR}^{-G(r,1)} \mathsf{SNR}^{-mn(1+G(r,1)-0)}\doteq \mathsf{SNR}^{-mn(1+G(r,1))-G(r,1)}$.
Further, we can use the same technique to find that for any $j<i$,
\begin{eqnarray}
\Pi(\jt=j|\jr=i)&\doteq& \mathsf{SNR}^{-mn(q_i-\max(q_j,0))}.
\end{eqnarray}
Thus we obtain the following result.


\begin{theorem}[Power-controlled Feedback]\label{thrmif} Suppose that $K>1$ and $r<\min(m,n)$. Then, the following diversity-multiplexing tradeoff can be achieved with power-controlled feedback
 \begin{eqnarray}
 &&\overline{d}_{\text{R}\widehat{\text{T}_\text{q}}}(r,K) =\min\left(d_{\text{R}{\text{T}}_q}(r,K),\max_{q_j\le1+d_{\text{R}{\text{T}}_q}(r,j)}\right.\nonumber\\
&&\left.\min_{i=1,\ldots,K-1}(mn((q_i)^+-(q_{i-1})^+)+d_{\text{R}{\text{T}}_q}(r,i)) \right).\nonumber
\end{eqnarray}
\end{theorem}
\begin{proof}
The proof is provided in Appendix \ref{apdxif}.
\end{proof}

\begin{corollary}
The diversity multiplexing tradeoff with one bit of imperfect feedback is same as the diversity multiplexing tradeoff with one bit of perfect feedback. In other words, for $K=2$, $\overline{d}_{\text{R}\widehat{\text{T}_\text{q}}}(r,2)=G(r,1+G(r,1))={d}_{\text{R}{\text{T}_\text{q}}}(r,2)$.
\end{corollary}
\begin{proof}
For $K=2$, the optimal choice of $q$ is to maximize $q_1^+-q_0^+$ which gives optimal choice of $q_0=0$ and $q_1=1+G(r,1)$. Using this for one bit of feedback, diversity of $G(r,1+G(r,1))$ can be achieved with power-controlled feedback which is the optimal considering the upper bound of perfect feedback is ${d}_{\text{R}{\text{T}_\text{q}}}(r,2)=G(r,1+G(r,1))$ .
\end{proof}
Thus, we find that there is no loss of diversity with one imperfect bit of feedback compared to the case of perfect feedback. Hence, power controlled feedback scheme is better than the constant power scheme which was limited to a maximum of $2mn$ diversity even as $r\to 0$.

\begin{corollary}
As $K\to\infty$ and $r\to 0$, the maximum diversity that can be obtained with power-controlled feedback is $mn(mn+2)$, i.e,
\[
\lim_{K \to \infty, r \to 0}\overline{d}_{\text{R}\widehat{\text{T}_\text{q}}}(r,K) = mn(mn+2).
\]
\end{corollary}
\begin{proof}
Note that in this case, choosing $q_0=0$ and $q_i= 1+d_{\text{R}{\text{T}}_q}(r,i) \approx 1+ mn\frac{mn^i-1}{mn-1}{\mathbf 1}_{mn>1}+i{\mathbf 1}_{mn=1}$ gives the optimal diversity multiplexing point for $r\to 0$ and thus, the diversity is $mn(mn+2)$. We further note that $K=3$ is enough to get this point.
\end{proof}

Thus, even with power-controlled feedback, arbitrary number of feedback bits do not yield unbounded increase in diversity order as in the case of CSIR$\text{T}_{\text{q}}$, where the diversity order increases unbounded with $K$.

Note that however, we restricted our attention to an ordering $q_j>q_{j-1}$ which can be relaxed giving better results as in the following Lemma. However, for our objective of the achievability for imperfect channel state at the receiver and imperfect channel state at the transmitter, the achievability strategy in Theorem \ref{thrmif} is enough. Further note that the new achievability strategy which relaxes the assumption of $q_j>q_{j-1}$ does give improved diversity, but still the diversity remain bounded with increase of the feedback levels.

\begin{lemma}
Let $q_i\ge 0$ and $q_i\ne q_j$ for any $i\ne j$, $0\le i,j\le K-1$. Then, the following diversity can be achieved by power-controlled feedback.

\begin{eqnarray}
&&\min\left(d_{\text{R}{\text{T}}_q}(r,K),
\max_{(q_0,\cdots,q_{K-1})\in Q}\right.\nonumber\\&&\left.\min_{i,j\in{0,\ldots,K-1}: j<i \text{ and } q_j<q_i}(mn(q_i-q_j)+d_{\text{R}{\text{T}}_q}(r,i)) \right),\nonumber
\end{eqnarray}
where $Q=\{(q_0, \cdots, q_{K-1}): q_j\le 1+\min(d_{\text{R}{\text{T}}_q}(r,j),\min_{i\in \{0,\cdots,j-1\}: q_i>q_j}(d_{\text{R}{\text{T}}_q}(r,i)+(q_i-q_j)mn)) \forall 0\le j \le K-1 \}$.
\end{lemma}
\begin{proof}
The proof is a simple generalization of the proof of Theorem \ref{thrmif}. The constraint on $Q$ keeps the $\Pi(\jt=i)\mathsf{SNR}^{q_i}\dot{\le} \mathsf{SNR}$ for the power constraint and the diversity expression records the possible dominant outage events.
\end{proof}

\subsection{A Source Coding Interpretation}

The idea of power-controlled feedback transmission is akin to source-coding the feedback information. In conventional lossless source coding (e.g. Huffman coding), the currency of representation is bits. There, to minimize average code-length, the codeword length is (approximately) inversely proportional to the probability of an event. Rare events are represented by longer codewords while more frequent events are represented by shorter length codewords, thereby minimizing the average codelength.

Analogously, our currency is average transmit power and objective is to minimize average error probability. Thus rare events get higher power and frequent events lower transmit power. Note that there are many power allocations which will meet the power constraint but they will all result in different error probabilities. Our proposed feedback transmit power allocation minimizes the error probability (in asymptotic sense).


\section{CSI$\widehat{\text{R}}\text{T}_{\text{q}}$: Estimated CSIR with Noiseless Quantized Feedback\label{imcs}}

In this section, we will consider the case when the receiver obtains its channel information from an MMSE estimate, $\widehat{H}$, using training. As a result, the receiver index $J_{\mathsf{R}}$ is not always equal to the optimal index, $J$, based on actual channel $H$. We will model the relationship between $J$ and $J_\mathsf{R}$ via an effective channel (described in Section~\ref{sec:power-controlled training}).
The feedback link, on the other hand, is assumed to be noiseless. As a result, the transmitter index $J_{\mathsf{T}} = J_{\mathsf{R}}$.

We will analyze two protocols. The first protocol, labeled constant-power training, trains the receiver once at the beginning using a constant power level. We show that feedback, even if noiseless, is completely useless in providing any gains in diversity order compared to a no-feedback system. Inspired by our understanding of the noisy feedback channel in Section~\ref{imfeed}, we propose a  second protocol, labeled \emph{power-controlled training} which utilizes the feedback and trains the receiver twice, where the second training is power-controlled based on feedback information. The second protocol has a higher diversity order than any non-feedback system.



\subsection{Training the receiver}
\label{sec:train}

In this subsection, we will consider MMSE channel estimation for a single user MIMO channel.
The channel is estimated using a training signal that is known at the receiver. From the received signal, MMSE estimation is done as in \cite{mmse} to get an estimate $\widehat{H}$ of the original channel $H$. Let $X_T$ be the training signal of size $m\times N$ for some $m\le N< T_{\rm coh}$ that is known at the receiver and transmitter.
The transmitter sends $X_T$ and the destination receives $Y_T=HX_T+W_T$ where $Y_T$ is a $n\times N$ received signal and $W_T$ is the additive Gaussian noise with each entry from $CN(0,1)$. Following \cite{mmse}, the optimal training signal is
\[
X_T = \left[\begin{array}{cc}\sqrt{(n\mu_0-1)^+}I_m & 0_{m\times(N-m)}\end{array}\right],
\]
where $\mu_0$ is tuned to satisfy power constraint, $I_m$ denote $m\times m$ identity matrix and $0_{m\times(N-m)}$ represents $m\times(N-m)$ matrix having all entries $0$. Hence, $n\mu_0-1 = \frac{N \mathsf{SNR}}{m}$. Further, the channel estimate is given by
\[\widehat{H}=Y_T(X_T^\dagger X_T+ I_N )^{-1}X_T^\dagger=Y_T\left[\begin{array}{c}\frac{\sqrt{\frac{N \mathsf{SNR}}{m}}}{1+\frac{N \mathsf{SNR}}{m}}I_m\\0_{N-m\times m}\end{array}\right],\]
which can be rewritten as
\begin{equation}\label{esti}\widehat{H}=H\frac{1}{1+\frac{m}{N \mathsf{SNR}}}+W_{2}\frac{\sqrt{\frac{N \mathsf{SNR}}{m}}}{1+\frac{N \mathsf{SNR}}{m}}\end{equation}
where $W_2$ is the left $n\times m$ submatrix of $W_T$. We now note some properties of MMSE estimate. First, it is easy to see that expected value of $(H-\widehat{H})\widehat{H}$ is zero confirming the orthogonality of the error with the unbiased estimate. Further, the variance of any entry in $(H-\widehat{H})$ is $\frac{1}{1+\frac{N \mathsf{SNR}}{m}}$. Thus, $H$ and $\widehat{H}$ are matrices in which each corresponding element is highly correlated with the correlation coefficient between corresponding element of $H$ and $\widehat{H}$ is $\rho={\frac{1}{\sqrt{1+\frac{m}{N \mathsf{SNR}}}}}$. Further note that any $N\ge m$ channel uses are equivalent for analyzing the asymptotic performance.

In general, if $G$ and $H$ are  correlated with correlation coefficient $\rho$, the joint probability distribution function of eigenvalues of $HH^\dagger$ and $GG^\dagger$ is given by the following result:

\begin{lemma}\cite{singular}\label{si} Consider two $n\times m$ random matrices $H = (h_{ij} )$
and $G = (g_{ij})$, $i \in [1,n]$, $j \in [1, m]$, each with i.i.d complex zero-mean unit-variance
Gaussian entries, i.e., $E[h_{ij} ] = E[g_{ij} ] = 0, \forall i, j, E[h_{ij}h^\dagger
_{pq}] = E[g_{ij}g^\dagger_{pq}] = \delta_{ip}\delta_{jq}$, where
the Kronecker symbol $\delta_{ij}$ is 1 or 0 when $i = j$ or $i \ne j$ respectively. Moreover, the correlation among the two random matrices is given by
$E[h_{ij}g^\dagger_{pq}] = \rho \delta_{ip}\delta_{jq}, \forall i, j, p, q,$ where $\rho = |\rho|e^{j\theta}$ is a complex number with $|\rho|<1$. Let $n\le m$ and $\nu=m-n$. The joint probability distribution function of the unordered eigenvalues of $HH^\dagger$ and $GG^\dagger$
is
\begin{eqnarray}
p(\lambda,\widehat{\lambda})&=&\frac{\exp\left(-\frac{\sum_{k=1}^n \lambda_k+\widehat{\lambda}_k}{1-|\rho|^2}\right)\triangle(\lambda)\triangle(\widehat{\lambda})}{n!n!\Pi_{j=0}^{n-1}j!(j+\nu)!|\rho|^{mn-n}(1-|\rho|^2)^n}\nonumber\\
&& \times \Pi_{k=1}^n (\sqrt{\lambda_k\widehat{\lambda}_k})^\nu\det\left|I_\nu\left(\frac{2|\rho|\sqrt{\lambda_k \widehat{\lambda}_l}}{1-|\rho|^2}\right)\right|,
\end{eqnarray}
where $\triangle(.)$ represents $n-$dimensional Vandermonde determinant, $I_k(.)$ denotes the $k^{th}$ order modified Bessel function of the first kind, the eigenvalues of $HH^\dagger$ and $GG^\dagger$ are given by $\lambda=(\lambda_1,\cdots,\lambda_n)$ and $\widehat{\lambda}=(\widehat{\lambda}_1,\cdots,\widehat{\lambda}_n)$ respectively.
\end{lemma}
Note that although Lemma \ref{si} assumed $n\le m$, it can be extended to the other case of $m>n$ since nonzero eigenvalues of $HH^\dagger$ and $H^\dagger H$ are the same. Hence for all $n$ and $m$, let $m_N=\min(m,n)$ and $\nu=|m-n|$. Then, the joint probability density function of the unordered eigenvalues of $HH^\dagger$ and $GG^\dagger$
is
\begin{eqnarray}\label{joint}
p(\lambda,\widehat{\lambda})&=&\frac{\exp\left(-\frac{\sum_{k=1}^{m_N} \lambda_k+\widehat{\lambda}_k}{1-|\rho|^2}\right)\triangle(\lambda)\triangle(\widehat{\lambda})}{m_N!m_N!\Pi_{j=0}^{n-1}j!(j+\nu)!|\rho|^{mn-m_N}}\nonumber\\
&& \times \frac{\Pi_{k=1}^{m_N} (\sqrt{\lambda_k\widehat{\lambda}_k})^\nu\det\left|I_\nu\left(\frac{2|\rho|\sqrt{\lambda_k \widehat{\lambda}_l}}{1-|\rho|^2}\right)\right|}{(1-|\rho|^2)^{m_N}} .
\end{eqnarray}

Recall that the eigenvalues of $HH^\dagger$ be $(\lambda_1, \cdots, \lambda_{m_N})$, $\lambda_i\doteq\mathsf{SNR}^{-\alpha_i}$ and ${\ba}=(\alpha_1,\cdots,\alpha_{m_N})$. Similarly, let the eigenvalues of $\widehat{H}\widehat{H}^\dagger$ be $(\widehat{\lambda}_1, \cdots, \widehat{\lambda}_{m_N})$, $\widehat{\lambda}_i\doteq\mathsf{SNR}^{-\widehat{\alpha}_i}$ and $\widehat{\ba}=(\widehat{\alpha}_1,\cdots,\widehat{\alpha}_{m_N})$.  The distribution of $\alpha_i$'s is given earlier in Lemma \ref{zh}. We will now find the joint distribution of $\alpha_i$'s and $\widehat\alpha_i$'s. Let $\alpha_1\ge \alpha_2\ge \cdots \alpha_{m_N}$ and $\widehat\alpha_1\ge \widehat\alpha_2\ge \cdots \widehat\alpha_{m_N}$. Further, we define
\begin{eqnarray}
E_{k}&=&\{({\ba,\bah}): \min(\alpha_i,\widehat{\alpha}_i)\ge1 \ \forall i=1,\cdots, k, \text{ and } \nonumber\\ &&\quad 0\le\alpha_i=\widehat{\alpha}_i<1 \ \forall i>k \}
\end{eqnarray}
 for all $0\le k\le m_N$.

\begin{theorem}\label{eigmimo} Let $H$ be the channel and $\widehat{H}$ be the estimated channel. In the limit of high $\mathsf{SNR}$, the probability density function of the $\mathsf{SNR}$
exponents of the eigenvalues of $HH^\dagger$ and $\widehat{H}\widehat{H}^\dagger$ is given by
\begin{eqnarray}
p({\ba,\bah})&\doteq& \sum_{k=0}^{m_N}e_{k}{\mathbf 1}_{E_{k}}
\end{eqnarray}
where $\alpha_1\ge \alpha_2\ge \cdots \alpha_{m_N}$, $\widehat\alpha_1\ge \widehat\alpha_2\ge \cdots \widehat\alpha_{m_N}$, and

\begin{eqnarray}
e_k&=&\mathsf{SNR}^{k(|n-m|+k)}\Pi_{i=1}^{k}\mathsf{SNR}^{-(2i-1+|n-m|)\widehat\alpha_i}\nonumber\\&&\times\Pi_{i=1}^{m_N}\mathsf{SNR}^{-(2i-1+|n-m|)\alpha_i}.
\end{eqnarray}
\end{theorem}
\begin{proof}
Since all the results were symmetric about interchanging $m$ and $n$, without loss of generality, we take $m\le n$ for the purpose of this proof and the rest of the paper. The proof is provided in Appendix \ref{jointapp}.
\end{proof}
\begin{remark}
Since the receiver is trained at $\mathsf{SNR}$, we note that for all $\alpha_i<  1$, $\alpha_i=\widehat{\alpha}_i$ with probability $1$, which means that the channel estimate is a reliable proxy for the actual channel. On the other hand, if $\alpha_i\ge 1$, all we can state is that $\widehat{\alpha}_i\ge 1$ with probability 1. For example, if $\widehat{\alpha}_i\ge 100$, all we can reliably say about $\alpha_i$ is that it is $\ge 1$. In SISO case, the above property of $\widehat{\alpha}_i$ implies that the channel cannot be resolved below the noise floor since the noise dominates the training signal. The interesting implication in MIMO is that this result of noise dominance holds for all the eigen-values. None of the eigen-value of the channel can be resolved beyond $\alpha_i\ge 1$ if $\widehat{\alpha}_i\ge 1$.
\end{remark}
\emph{Example 4 (Asymptotic Distribution)}: For the case of $m=n=1$, which is our running example, $p({\alpha,\widehat{\alpha}})\doteq \mathsf{SNR}^{-\alpha}{\mathbf 1}_{0\le\alpha=\widehat{\alpha}<1}+\mathsf{SNR}^{1-\alpha-\widehat{\alpha}}{\mathbf 1}_{\min(\alpha,\widehat{\alpha})\ge1}$.
This density function will be used to analyze the diversity tradeoffs in this section.

\subsection{Constant-power Training \label{sec:constant-power training}}


We first note the decoding scheme with trained channel estimate at the receiver. If the receiver has estimate $\widehat{H}$ trained with the power of $\mathsf{SNR}^p$ for some $p\ge 0$, the estimation error variance is $\mathsf{SNR}^{-p}$. As $Y=HX+W = \widehat{H}X+(H-\widehat{H})X+W$, a lower bound on mutual information can be considered assuming that $(H-\widehat{H})X+W$ is Gaussian noise and hence the expression for the coherent channel with actual channel as $\widehat{H}$ can be used as a lower bound \cite{zhengnon}.

The protocol is divided into three phases as described below:

\noindent  {\bf Phase 1:} Training from Tx to Rx: The training is done using transmit power of $\mathsf{SNR}$ to obtain the channel estimate $\widehat{H}$. On the basis of this training, the receiver decides a feedback level $\jr$ in the following way.  Suppose that there are $K$ power levels $P_i\doteq \mathsf{SNR}^{1+p_i}, i\in\{0,\cdots,K-1\}$. The feedback index chosen at the receiver is given by
\[
\jr =
\begin{cases}
\arg \min_{i \in \mathbb{I}}  i,  & \mathbb{I} = \{ k : \log \det(I+\widehat{H}\widehat{H}^\dagger {P_{k}})\ge R, \\
& \quad k \in \{0,\cdots,K-1\} \}  \\
  K-1,  & \text{ if the set $\mathbb{I}$ is empty } \\
\end{cases}.
\]

\noindent  {\bf Phase 2:}  The feedback index $\jr\in\{0,\cdots,K-1\}$ is sent to the transmitter via the noiseless feedback channel. Thus, the feedback index received by the transmitter is $\jt=\jr$.

\noindent  {\bf Phase 3:}  The transmitter receives a feedback power level $\jt$ and sends data at power level $P_{\jt}\doteq \mathsf{SNR}^{1+p_{\jt}}$.

\noindent
\emph{Example 5 (Constant-power training)}: The only error in channel knowledge appears in the first phase, where the receiver is trained.

Note that for MMSE estimation $\widehat{H}$ and $\widetilde{H}=H-\widehat{H}$ are uncorrelated and since the receiver is trained with power of $\mathsf{SNR}$, the variance of $\widetilde{H}$ is $\doteq 1/\mathsf{SNR}$. Let $|\widehat{H}|^2=\mathsf{SNR}^{-\widehat{\alpha}}$ and $|\widetilde{H}|^2=\mathsf{SNR}^{-1}\mathsf{SNR}^{-\widetilde{\alpha}}$ where the probability distribution function of $\widehat{\alpha}$ and $\widetilde{\alpha}$ is $p(\widehat{\alpha},\widetilde{\alpha})=\mathsf{SNR}^{-\widehat{\alpha}-\widetilde{\alpha}}{\mathbf 1}_{\widehat{\alpha}\ge 0}{\mathbf 1}_{\widetilde{\alpha}\ge 0}$. The probability of outage is
\begin{eqnarray}
\Pi({\cal O})&=&\sum_{j=0}^{K-1}\Pi({\cal O},\jt=j)\nonumber\\
&\ge&\Pi({\cal O},\jt=1)\nonumber\\
&=&\Pi({\cal O},\jr=1)\nonumber\\
&\dot{\ge}&\Pi({\cal O},\log(1+|\widehat{H}|^2\mathsf{SNR}^2)\ge r\log\mathsf{SNR},\nonumber\\
&&\quad \log(1+|\widehat{H}|^2\mathsf{SNR})< r\log\mathsf{SNR})\nonumber\\
&\doteq&\Pi\left(\log\left(1+\frac{|\widehat{H}|^2\mathsf{SNR}^2}{1+\mathsf{SNR}^2|\widetilde{H}|^2}\right)<r\log\mathsf{SNR},\right.\nonumber\\
&&\quad \log(1+|\widehat{H}|^2\mathsf{SNR}^2)\ge r\log\mathsf{SNR},\nonumber\\
&&\left.\quad \log(1+|\widehat{H}|^2\mathsf{SNR})< r\log\mathsf{SNR}\right)\nonumber\\
&\doteq&\Pi((2-\widehat{\alpha}-(1-\widetilde{\alpha})^+)^+<r,(2-\widehat{\alpha})^+\ge r,\nonumber\\
&&\quad (1-\widehat{\alpha})^+< r)\nonumber\\
&\doteq&\mathsf{SNR}^{-\min_{(\widehat{\alpha},\widetilde{\alpha})\in A}(\widehat{\alpha}+\widetilde{\alpha})},
\end{eqnarray}
where $A=\{(\widehat{\alpha},\widetilde{\alpha}): \widehat{\alpha}\ge 0,\widetilde{\alpha}\ge 0,(2-\widehat{\alpha}-(1-\widetilde{\alpha})^+)^+<r,(2-\widehat{\alpha})^+\ge r,(1-\widehat{\alpha})^+< r\}$.

Hence, substituting $\widetilde{\alpha}=0$ and $\widehat{\alpha}=1-r+\delta$ for $\delta$ small enough (Note that this choice of $(\widehat{\alpha},\widetilde{\alpha})$ is in $A$.) gives a bound on the above probability as:
\begin{eqnarray}
\Pi({\cal O})&\dot{\ge}&\mathsf{SNR}^{-\min_{(\widehat{\alpha},\widetilde{\alpha})\in A}(\widehat{\alpha}+\widetilde{\alpha})}\nonumber\\
&\dot{\ge}&\mathsf{SNR}^{-(1-r)}.
\end{eqnarray}
Thus, we see that the diversity of this scheme is at most same as the one without feedback. Hence, there is no advantage of feedback.\hfill $\blacksquare$

\begin{theorem} Suppose that $K>1$ and $r<\min(m,n)$. Then, the diversity-multiplexing tradeoff is given by $d_{\widehat{\text{R}}\text{T}_q}(r,K) = G(r,1)$.
\end{theorem}
\begin{proof} We will consider the third phase in this case, where we see that any increase in power levels do not help to increase the diversity. Achievability follows by not using any feedback. We will prove the converse here.
Note that for MMSE estimation $\widehat{H}$ and $\widetilde{H}=H-\widehat{H}$ are uncorrelated and since the receiver is trained with power of $\mathsf{SNR}$, the variance of $\widetilde{H}$ is $\doteq 1/\mathsf{SNR}$. Let $\widehat{\lambda}_i$ be the eigenvalues of $\widehat{H}\widehat{H}^\dagger$ and $\widehat{\lambda}_i\doteq \mathsf{SNR}^{-\widehat{\alpha}_i}$. Further, assume that $\widetilde{\lambda}_i$ be the eigenvalues of $\widetilde{H}\widetilde{H}^\dagger$ and $\widetilde{\lambda}_i\doteq \mathsf{SNR}^{-1}\mathsf{SNR}^{-\widetilde{\alpha}_i}$. The probability distribution function of $\widehat{\ba}=(\widehat{\alpha}_1,\cdots,\widehat{\alpha}_m)$ and $\widetilde{\ba}=(\widetilde{\alpha}_1,\cdots,\widetilde{\alpha}_m)$ is $p(\widehat{\ba},\widetilde{\ba})=\Pi_{i=1}^{m_N}\mathsf{SNR}^{-(2i-1+|n-m|)(\widehat{\alpha}_i+\widetilde{\alpha}_i)}{\mathbf 1}_{\min(\widehat{\ba})\ge0}{\mathbf 1}_{\min(\widetilde{\ba})\ge0}$. The probability of outage is
\begin{eqnarray}
\Pi({\cal O})&=&\sum_{j=0}^{K-1}\Pi({\cal O},\jt=j)\nonumber\\
&\ge&\Pi({\cal O},\jt=1)\nonumber\\
&=&\Pi({\cal O},\jr=1)\nonumber\\
&\dot{\ge}&\Pi({\cal O},\log\det(I+\widehat{H}\widehat{H}^\dagger P_1)\ge r\log\mathsf{SNR},\nonumber\\
&&\quad \log\det(I+\widehat{H}\widehat{H}^\dagger P_0)< r\log\mathsf{SNR})\nonumber\\
&\doteq&\Pi\left(\log\det\left(I+\frac{\widehat{H}\widehat{H}^\dagger P_1}{1+P_1 {\rm trace}(\widetilde{H}\widetilde{H}^\dagger)}\right)\right.\nonumber\\
&&\quad <r\log\mathsf{SNR}, \nonumber\\
&&\quad \log\det(I+\widehat{H}\widehat{H}^\dagger P_1)\ge r\log\mathsf{SNR},\nonumber\\
&&\quad \left.\log\det(I+\widehat{H}\widehat{H}^\dagger P_0)< r\log\mathsf{SNR}\right)\nonumber\\
&\doteq&\Pi\left(\sum_{i=1}^m(1+p_1-\widehat{\alpha}_i-(p_1-\min\widetilde{\ba})^+)^+<r,\right.\nonumber\\&&\quad\sum_{i=1}^m(1+p_1-\widehat{\alpha}_i)^+\ge r,\nonumber\\
&&\left.\quad \sum_{i=1}^m(1+p_0-\widehat{\alpha}_i)^+< r\right)\nonumber\\
&\doteq&\mathsf{SNR}^{-\min_{(\widehat{\ba},\widetilde{\ba})\in A}\sum_{i=1}^m((2i-1+|n-m|)(\widehat{\alpha}_i+\widetilde{\alpha}_i))}
\end{eqnarray}
where $A=\{(\widehat{\ba},\widetilde{\ba}): \min\widehat{\ba}\ge 0,\widetilde{\ba}\ge 0,\sum_{i=1}^m(1+p_1-\widehat{\alpha}_i-(p_1-\min\widetilde{\ba})^+)^+<r,\sum_{i=1}^m(1+p_1-\widehat{\alpha}_i)^+\ge r,\sum_{i=1}^m(1+p_0-\widehat{\alpha}_i)^+< r\}$.
Hence, substituting $\widetilde{\alpha}_i=0$ and allocating $\widehat{\alpha}$ as in \cite{kim07} so that $\sum_{i=1}^m(1+p_0-\widehat{\alpha}_i)^+= r-\delta$  for $\delta$ small enough gives a bound on the above probability as:
\begin{eqnarray}
\Pi({\cal O})&\dot{\ge}&\mathsf{SNR}^{-\min_{(\widehat{\ba},\widetilde{\ba})\in A}\sum_{i=1}^m((2i-1+|n-m|)(\widehat{\alpha}_i+\widetilde{\alpha}_i))}\nonumber\\
&\dot{\ge}&\mathsf{SNR}^{-G(r,1+p_0)}\nonumber\\
&\dot{\ge}&\mathsf{SNR}^{-G(r,1)}.
\end{eqnarray}
In the last step, $p_0\le 0$ else the transmit power constraint cannot be satisfied. Thus, we see that the diversity of this scheme is at most same as the one without feedback. Hence, there is no advantage of feedback for constant power training. This was observed for $r\to 0$ in \cite{gkthesis}.
\end{proof}
\begin{remark}
Analogous to Remark 3, the accounting for the training resources can be done. In the limit of high $\mathsf{SNR}$, the training will consume $m$ channel uses to train $m$ antennas (as also seen in Section \ref{sec:train}) and hence to get the rate $R\doteq r\log(\snr)$, $r$ should be replaced by $rT_{\rm coh}/(T_{\rm coh}-m)$ in the above expression. However, these terms will be omitted in the sequel and can be similarly integrated. Much like in Remark 3, one may carry out an optimization on the diversity obtained as a function of these time loss terms for each multiplexing to optimize over the number of antennas that need to be trained like in \cite{zhengnon,steger}.
\end{remark}

%
%
%
%

\subsection{Power-controlled Training \label{sec:power-controlled training}}

As we showed in the previous section, the receiver estimate was not good enough to help improve the diversity with feedback. We now propose a power-controlled training protocol which can improve the outage performance using feedback.
The protocol is again divided into three phases as described below:

\noindent  {\bf Phase 1:}  The transmitter sends the training signal using power $\mathsf{SNR}$ which is used at the receiver to obtain channel estimate $\widehat{H}$. On the basis of this training, the receiver decides a feedback level $\jr$ in the following way. Suppose that there are $K$ power levels $P_i\doteq \mathsf{SNR}^{1+p_i}, i\in\{0,\cdots,K-1\}$. (We will give the exact constants in front while proving achievability.) The feedback index is
\[
\jr =
\begin{cases}
\arg \min_{i \in \mathbb{I}}  i,  & \mathbb{I} = \{ k : \log \det(I+\widehat{H}\widehat{H}^\dagger {P_{k}})\ge \\
&\quad R+\epsilon\log(\snr),\nonumber\\
&\quad k \in \{0,\cdots,K-1\} \}  \\
  K-1,  & \text{ if the set $\mathbb{I}$ is empty } \\
\end{cases},
\]
where $\epsilon>0$ is some small constant chosen. We will later substitute $\epsilon\to 0$.

\noindent  {\bf Phase 2:}  A feedback level $\jr\in\{0,\cdots,K-1\}$ is fed back over the noiseless feedback channel to the transmitter, which implies $\jt=\jr$.

\noindent  {\bf Phase 3:}  If the feedback index $\jt$ is received, the transmitter trains the receiver \emph{again} at power level $P_{\jt}$ which is followed by the data at power level $P_{\jt}$. The trained channel estimate is denoted by $\widehat{H}_2$ and let $\widetilde{H}=H-\widehat{H}_2$.

The outage probability is defined as: $\Pi({\cal O})\triangleq$

\begin{equation}\label{outexpr}\Pi\left(\log \det\left(I+\frac{P(\jt)}{m}\frac{\widehat{H}_2 Q\widehat{H}_2^\dagger}{1+\frac{\mathsf{SNR}}{mn}{\rm trace}(\widetilde{H}\widetilde{H}^\dagger)}\right)<R \right)
\end{equation}
This is the effective outage probability using Gaussian codebooks and considering the channel estimation error as noise \cite{steger}.

\begin{figure*}[htbp]
\centering \myincludegraphics[width=6.5in]
{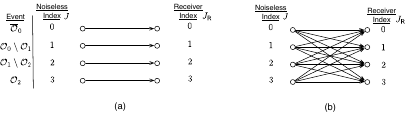}\caption{Example 6: Obtaining channel information at the receiver: (a)~if the receiver has perfect information, then it knows the correct feedback index, else (b)~the receiver index is not known perfectly and can be viewed as an output of a noisy channel.}\label{fig:noisy-CSIR}
\end{figure*}

\emph{Example 6 (Power-controlled Training)}: Since the receiver now does not know the value of channel estimate $H$, but only an estimate $\widehat{H}$, the events $\widehat{\cal O}_i$ can in this case be defined as
\begin{equation}
\widehat{\cal O}_i = \left\{ \widehat{H} : \log\left(1+ |\widehat{H}|^2 \mathsf{SNR}^{1+i} \right) < R \right\}, i = 0,1,2.
\end{equation}
Figure~\ref{fig:noisy-CSIR} depicts the relation between $J$ and $J_{\mathsf R}$.
%
%
Let $\alpha$ be the negative $\mathsf{SNR}$ exponent of $|H|^2$ while $\widehat{\alpha}$ be the negative $\mathsf{SNR}$ exponent of $|\widehat{H}|^2$. Let

\[
J =
\begin{cases}
\arg \min_{i \in \mathbb{I}}  i,  & \mathbb{I} = \{ k : \log (1+|H|^2 {P_{k}})\ge R, \\
& \quad k \in \{0,\cdots,3\} \}  \\
  0,  & \text{ if the set $\mathbb{I}$ is empty } \\
\end{cases}.
\]

 We assume that  $3^{rd}$ phase is perfect for this example and hence $\widetilde{H}=0$. We will prove later that this interference error in \eqref{outexpr} due to $\widetilde{H}$ does not make a difference asymptotically.

Consider the event $(J_{\mathsf R}<2,J=2)$ which would result in outage. The powers and the probabilities can be seen in Table \ref{tbl:imcsir}. The outage probability for $r\to 0$ is then
\begin{eqnarray}
\Pi({\cal O})&\ge&\Pi(J_{\mathsf R}<2,J=2)\nonumber\\
&=&\Pi(\log \det(I+\widehat{H}\widehat{H}^\dagger {P_{1}})\ge R,\nonumber\\
&&\quad \log (1+|H|^2 {P_{1}})< R,\nonumber\\
&&\quad \log (1+|H|^2 {P_{2}})\ge R)\nonumber\\
&\ge&\Pi(2-\widehat{\alpha}>0,2-{\alpha}<0,3-{\alpha}>0)\nonumber\\
&\doteq&\mathsf{SNR}^{\max_{2<\alpha<3,1<\widehat{\alpha}<2}(1-\alpha-\widehat{\alpha})}\nonumber\\
&\doteq&\mathsf{SNR}^{1-2-1}\doteq \mathsf{SNR}^{-2}
\end{eqnarray}
Thus, the maximum diversity order is $2$ with any number of feedback levels. Note that this is more than constant power training which was limited to $1$. We will show later in this Section that this can be achieved with a single feedback bit.\hfill $\blacksquare$
\begin{table}
\begin{center}
\caption{Example 6: Power assignment for CSI$\widehat{\text{R}}\text{T}_{\text{q}}$.  (Caution: Probabilities are only reported up to their order, and constants such that they sum to one are omitted.) \label{tbl:imcsir}}
\begin{tabular}{|c|c|c|c|} \hline
Event & Prob at Receiver &  Training and\\
& & Transmit Power   \\ \hline
$\overline{\widehat{\cal O}}_0$ & $1-\mathsf{SNR}^{-1}$  & $\mathsf{SNR}^{1}$  \\
$\widehat{\cal O}_0\setminus {\cal O}_1$ & $\mathsf{SNR}^{-1}$ &  $\mathsf{SNR}^{2}$  \\
$\widehat{\cal O}_1\setminus {\cal O}_2$ & $\mathsf{SNR}^{-2}$ &  $\mathsf{SNR}^{3}$  \\
$\widehat{\cal O}_2$ & $\mathsf{SNR}^{-3}$ &  $\mathsf{SNR}^{4}$ \\ \hline
\end{tabular}
\end{center}
\end{table}

The above example can be generalized to MIMO systems as follows.
\begin{theorem}\label{thrm:csihatrt}
For $K>1$ and $r<\min(m,n)$, the diversity-multiplexing tradeoff of ${\overline d}_{\widehat{\text{R}}\text{T}_q}(r,K)=G(r,1+G(r,1))$ can be achieved with power-controlled training. Further, the above is optimal for zero multiplexing.
\end{theorem}
\begin{proof}
The proof of this Theorem is provided in Appendix \ref{app:csihatrt}. We will first show that the diversity cannot be greater than $mn(mn+1)$ and later prove that the diversity multiplexing tradeoff of $G(r,1+G(r,1))$ can be achieved for $K=2$.
\end{proof}

Note that ${\overline d}_{\widehat{\text{R}}\text{T}_q}(r,K)=d_{\text{RT}_q}(r,2)$ which means that diversity with imperfect receiver information with any number of feedback levels $K\ge 2$ is same as the diversity with perfect receiver information with $1$ bit of feedback.

\section{CSI$\widehat{\text{R}}\widehat{\text{T}}_{\text{q}}$: Estimated CSIR with Noisy Quantized Feedback}\label{mainsec}

We observed in Section \ref{imcs} that with CSIR obtained by MMSE training and perfect feedback,
the diversity-multiplexing tradeoff of ${\overline d}_{\widehat{\text{R}}\text{T}_q}(r,K)=G(r,1+G(r,1))$
can be achieved with 1 bit of noiseless feedback. We also observed that the diversity-multiplexing tradeoff of
  ${\overline d}_{\text{R}\widehat{\text{T}}_q}(r,K)=G(r,1+G(r,1))$ is also achievable with 1 bit of
  noisy feedback with perfect CSIR. In this section, we will show that diversity-multiplexing tradeoff
  ${\overline d}_{\widehat{\text{R}}\text{T}_q}(r,2)$ can be achieved when both imperfections are present
  simultaneously: 1 bit of imperfect feedback based on noisy training-based receiver information.  Thus, our main result is
\begin{theorem}
For $K=2$ and $r<\min(m,n)$, an achievable diversity-multiplexing tradeoff is given by ${\overline d}_{\widehat{\text{R}}\widehat{\text{T}}_q}(r,2)=G(r,1+G(r,1))$.
\label{th:main-result}
\end{theorem}

%

\subsection{Protocol}
The complete protocol with both power-controlled training and power-controlled feedback constitutes of three phases as described below.

\noindent  {\bf Phase 1:}  The training is done using power $\mathsf{SNR}$ to get the channel estimate $\widehat{H}$. On the basis of this training, the receiver decides a feedback level $\jr$ in the following way. Since the feedback is assumed to be only one bit, there are only two power levels at the transmitter. Denote the two power levels as $P_i\doteq \mathsf{SNR}^{1+p_i}, i\in\{0,1\}$ with $p_0=0$ and $p_1=G(r,1)$. (We will give the exact constants in front while proving that the average power constraint will be satisfied.) The feedback index is
\[\jr= \begin{cases}
  1 &\text{ if } \log \det\left(I+\widehat{H}\widehat{H}^\dagger {P_{0}}\right)<R+\epsilon\log(\snr)\\
  0  &\text{ otherwise } \\
 \end{cases},
 \]
 where $\epsilon>0$ is an arbitrarily small constant chosen as before.
The above index assignment is simply choosing the higher power level if the lower power level is estimated to be too low.

\noindent  {\bf Phase 2:}  A feedback level $\jr\in\{0,1\}$ is transmitted from the receiver which is received at the transmitter as  $\jt\in\{0,1\}$. The receiver employs the power-controlled encoding scheme described in Section \ref{imfeed} to send the feedback index. The power levels used for sending the feedback information are denoted as $Q_0=0$ and $Q_1\doteq \mathsf{SNR}^{1+G(r,1)}$.

\noindent  {\bf Phase 3:}  The transmitter gets a feedback index $\jt$ which is used to train the receiver at power level $P_{\jt}$ and then send data at power level $P_{\jt}$. The trained channel estimate is denoted $\widehat{H}_2$ and let $\widetilde{H}=H-\widehat{H}_2$.

The outage probability of the above three-phase protocol is upper bounded by
\begin{eqnarray}
\Pi({\cal O})&=&\Pi\left(\log \det\left(I+\frac{P_{\jt}}{m}\right. \right.\nonumber\\
&&\left. \left. \quad \frac{\widehat{H}_2 Q\widehat{H}_2^\dagger}{1+\frac{P_{\jt}}{mn}{\rm trace}(\widetilde{H}\widetilde{H}^\dagger)}\right)<R\right).
\end{eqnarray}
The above expression for the effective outage probability considers the channel estimation error as noise~\cite{steger} and hence is only an upper bound for the optimal scheme. Following identical steps as in Section \ref{imcs}, we find that the interference error due to $\widetilde{H}$ does not impact the analysis asymptotically and can thus be ignored. Hence,
\begin{equation}
\Pi({\cal O})\doteq\Pi\left(\log \det\left(I+P(\jt)\widehat{H}_2\widehat{H}_2^\dagger\right)<R\right). \label{eq:effective-outage}
\end{equation}
Now define $J$ as
\[
{J}= \begin{cases}
  1  &\text{ if } \log \det\left(I+\widehat{H}_2\widehat{H}_2^\dagger P_0\right)<R \text{ and }\\
   & \quad \log \det\left(I+\widehat{H}_2\widehat{H}_2^\dagger P_1\right)\ge R\\
  0  &\text{ otherwise } \\
 \end{cases}.
 \]
Using the analysis as in Sections III and IV, we observe that
\begin{eqnarray*}
\Pi(\jt=0|\jr=1) &\doteq& \mathsf{SNR}^{-mn(1+G(r,1))}, \\
\Pi(\jt=1|\jr=0) &\doteq& 0, \\
\Pi({\jr}=0,{J}=1) &\doteq& 0.
\end{eqnarray*}

\noindent We now split $\Pi({\cal O})$ in~(\ref{eq:effective-outage}) into 8 terms depending on the values of $J$, $\jr$ and $\jt$ as follows,
\begin{eqnarray}
\Pi({\cal O})&\doteq&\Pi(\log \det(I+P(\jt)\widehat{H}_2\widehat{H}_2^\dagger)<R)\nonumber\\
&\doteq&\sum_{i=0}^1\sum_{j=0}^1\sum_{k=0}^1\Pi(\log \det(I+P(\jt) \widehat{H}_2\widehat{H}_2^\dagger)<R,\nonumber\\
&& \quad J=i, \jr=j, \jt=k).
\end{eqnarray}
Note that the terms with  $\jr \neq \jt$ can be asymptotically upper bounded $\mathsf{SNR}^{-G(r,1+G(r,1))}$. Also, the term corresponding to $i=j=k=0$ can be upper bounded by $\mathsf{SNR}^{-G(r,1+G(r,1))}$ since when $J=0$, $\log \det(I+P(0)\widehat{H}_2\widehat{H}_2^\dagger)<R$ happens when $\log \det(I+P_1\widehat{H}_2\widehat{H}_2^\dagger)<R$. Further, $\jr=0$ and $J=1$ happens with probability $\doteq 0$. Thus, the only remaining case is when $\jt=\jr=1$.
 Hence,
\begin{eqnarray}
\Pi({\cal O})&\doteq&\sum_{i=0}^1\sum_{j=0}^1\sum_{k=0}^1\Pi(\log \det(I+P(\jt)\widehat{H}_2\widehat{H}_2^\dagger)<R,\nonumber\\
&& \quad J=i, \jr=j, \jt=k)\nonumber\\
&\dot{\le}&\mathsf{SNR}^{-G(r,1+G(r,1))}\nonumber\\
&&+\Pi(\log \det(I+P(\jt)\widehat{H}_2\widehat{H}_2^\dagger)<R,\nonumber\\
&& \quad \jt=1,\jr=1)\nonumber\\
&\dot{\le}&\mathsf{SNR}^{-G(r,1+G(r,1))}\nonumber\\
&&+\Pi(\log \det(I+P_1\widehat{H}_2\widehat{H}_2^\dagger)<R\nonumber\\
&\doteq&\mathsf{SNR}^{-G(r,1+G(r,1))}.
\end{eqnarray}
Hence, the diversity order of $G(r,1+G(r,1))$ for multiplexing gain $r$ can be achieved with only one bit of feedback. We now show that the power constraint is also satisfied which completes the proof. Recall that the power levels at the transmitter are denoted by $P_0$ and $P_1$, while at the receiver are denoted by $Q_0$ and $Q_1$.

Let $P_0=\frac{\mathsf{SNR}}{2}$ and $P_1 =
\frac{\mathsf{SNR}}{4(\Pi(\jr=1))}$. The average power used at the transmitter is $P_0 \Pi(\jt=0)+P_1 \Pi(\jt=1)$.
To show that this average power $\le \mathsf{SNR}$, it is enough to prove that
\begin{equation}\Pi(\jt=1)\le 2(\Pi(\jr=1)).\end{equation} Thus, the left hand side
\begin{eqnarray}
\Pi(\jt=1)&=&\Pi(\jt=1,\jr=0)+\Pi(\jt=1,\jr=1)\nonumber\\
&\le& \Pi(\jt=1|\jr=0)+\Pi(\jr=1).
\end{eqnarray}
Note that the first decays faster with $\mathsf{SNR}$ than $\Pi(\jr=1)$, this the above $\le 2\pi(\jr=1)$.

Further, let $Q_0=0$ and $Q_1 =
\frac{\mathsf{SNR}}{2(\Pi(\jr=1))}$. The average power used at the receiver is $Q_0 \Pi(\jr=0)+Q_1 \Pi(\jr=1)\le \snr$.


Thus, diversity order of $G(r,1+G(r,1))$ can be achieved with imperfect feedback and imperfect CSIR. Hence, one bit of imperfect feedback and imperfect CSIR is same as one bit of perfect feedback and perfect CSIR except the time losses in trainings and the feedback.

\section{Numerical Results}

\begin{figure}[htbp]
\begin{center} \includegraphics[trim=3.5cm 7.7cm 3.9cm 7.7cm, clip,width=3.3in]
{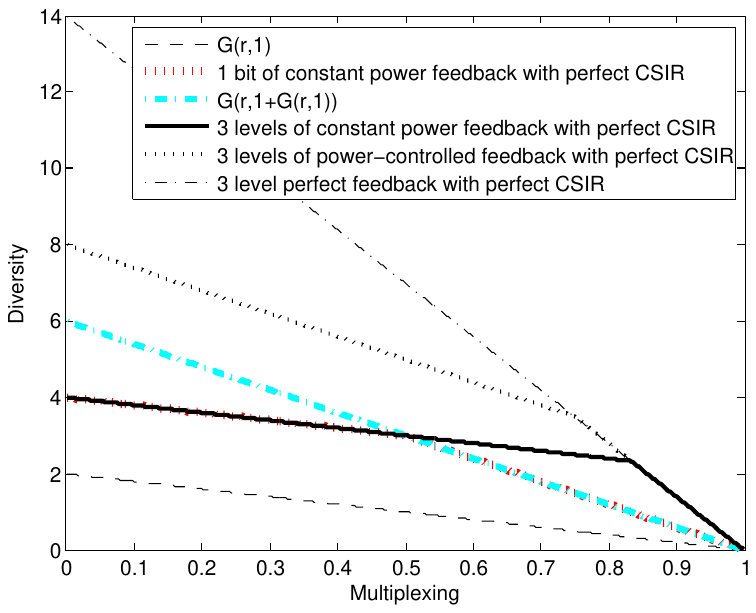}\caption{Diversity multiplexing tradeoff in various scenarios.}\label{fig_dmm1n2}
\end{center}
\end{figure}

First consider the case of $m=1$ and $n=2$. The diversity multiplexing tradeoff in the various cases for 1 or 2 bits of feedback can be seen in Figure \ref{fig_dmm1n2}. We will now go through all the different tradeoff curves in the order of the legend from top to bottom. The first line $G(r,1)$ represents the diversity obtained with no feedback (CSIR and CSI$\widehat{\text{R}}$ have identical performance since time lost in training is not accounted in our expressions) and also the diversity obtained in the case when the feedback is perfect and the receiver is trained through a constant power symbol (CSI$\widehat{\text{R}}\text{T}_\text{q}$ with constant power training). The second piecewise linear curve represents the diversity obtained when the receiver knows perfect channel state information but one bit of constant power feedback is sent over a noisy feedback channel (CSIR$\widehat{\text{T}}_\text{q}$ with constant power feedback). The third line $G(r,1+G(r,1))$ represents the diversity obtained with $1$ bit of perfect feedback (CSIR$\text{T}_\text{q}$). This also represents the diversity obtained when receiver knows the channel perfectly while $1$ bit of power-controlled feedback is provided on a feedback channel (CSIR$\widehat{\text{T}}_\text{q}$ with power-controlled feedback). This diversity is also obtained when the feedback link is perfect while the receiver is trained using power-controlled training symbols (CSI$\widehat{\text{R}}\text{T}_\text{q}$ with power-controlled training). Further, the diversity obtained when the receiver does not know channel state information and the feedback link is noisy is also $G(r,1+G(r,1))$ (CSI$\widehat{\text{R}}\widehat{\text{T}}_\text{q}$ with power-controlled training and power-controlled feedback). Note that the second curve started from diversity of $4$ at zero multiplexing, but then followed the third line after $r\approx 0.5$ since it cannot perform better than the perfect feedback case.

Next, we consider three levels of feedback. More bits of feedback do not increase the diversity in any case except when the receiver knows the channel perfectly. In this case, the next two curves lines show the effect of constant power feedback and the power controlled feedback on the diversity. The last line shows the performance with $3$ levels of perfect feedback. The diversity multiplexing tradeoff curve achieved with constant power starts from $4$ and hits the line of $3$ levels of perfect feedback.

\begin{figure}[htbp]
\begin{center} \includegraphics[trim=3.5cm 7.7cm 3.9cm 7.7cm, clip,width=3in]
{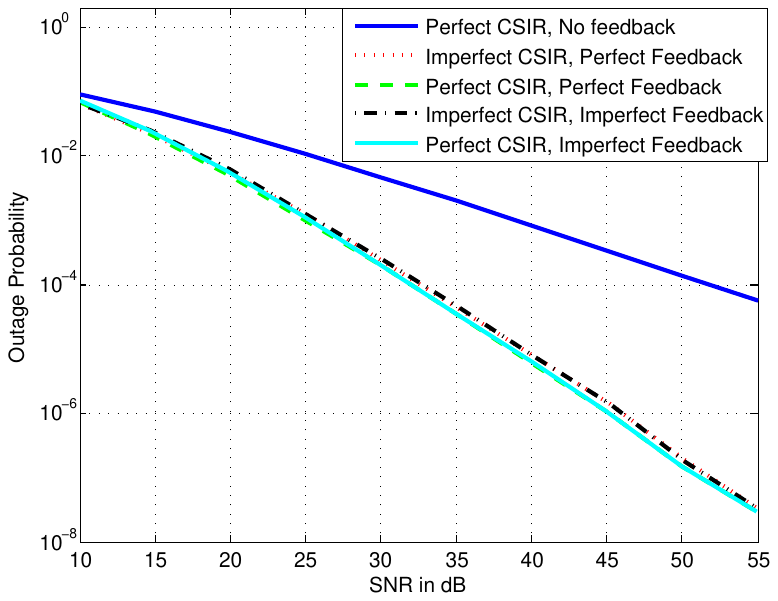}\caption{Diversity multiplexing tradeoff in various scenarios. The various parameters chosen are: m=1, n=1, r=0.2. The channel estimates are obtained using a training time of 10 channel uses.}\label{fig_m1n1}
\end{center}
\end{figure}

\begin{figure}[htbp]
\begin{center} \includegraphics[trim=4cm 8cm 4.5cm 8cm, clip,width=3in]
{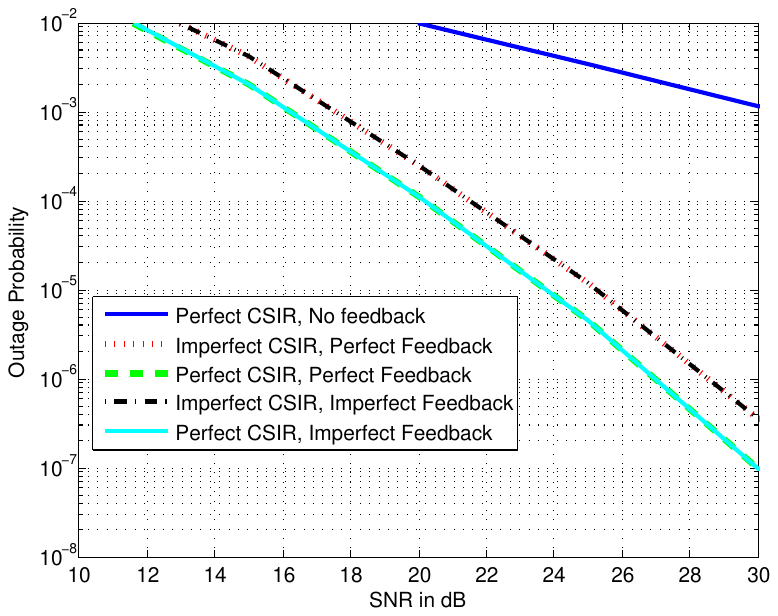}\caption{Diversity multiplexing tradeoff in various scenarios. The various parameters chosen are: m=1, n=2, r=0.5.  The channel estimates are obtained using a training time of 10 channel uses.}\label{fig_m1n2}
\end{center}
\end{figure}

We will now see the effect of increase on $\mathsf{SNR}$ on the outage probabilities at a constant multiplexing gain. We will focus on $1$ bit of feedback. In Figure \ref{fig_m1n1}, $m=n=1$ and $r=0.2$. Thus, the theoretical diversity for the case of no feedback is $0.8$ while for all other cases considered in the Figure is $1.6$ and we find that we obtain close to the expected diversity order at $\mathsf{SNR}$ of about $20$ dB. Note that the differences in the higher diversity order curves is small and they seem on top of each other in the plot.
In Figure \ref{fig_m1n2}, $m=1$, $n=2$ and $r=0.5$. In this case, the theoretical diversity for the case of no feedback is $1$ while with feedback is $3$ which can be noted from the slope.


\section{Extension to Multiuser MIMO \label{sec:multiple_access}}

In this section, we extend our main result for point-to-point systems, i.e, Theorem~\ref{th:main-result}, to the case of multiple access channel under the model of common feedback to all transmitters.

\subsection{Multiuser Channel Model}\label{sec:mac_channel model}

Consider a multiple access channel with $L$ transmitters ${\mathsf T}_i$ where each
transmitter has  $m$ transmit antennas and the receiver ${\mathsf R}$
has $n$ receive antennas. The channel is constant during
a fading block of $T_{\rm coh}$ channel uses, but changes independently from
one block to the next.

The received signal can be written in the matrix form as
\begin{equation}
\{ \mathsf{T}_i \} \rightarrow \mathsf{R}: Y = \mathop \sum \limits_{1 \le i \le L} H_{i}X_{i}+ W.
\end{equation}
Here, $W$ is of size $n \times T_{\rm coh}$ represents additive white Gaussian
noise at the receiver with all entries i.i.d. $CN(0,1)$. We consider
a Rayleigh fading environment, i.e. elements of
$H_{i}$ are assumed to be i.i.d $CN(0,1)$. The transmitters are
subject to an average power constraint such that the long-term power
is upper bounded, i.e, $\frac{1}{T_{\rm coh}}{\rm trace}({\mathbb E}\left[ X_{i}X_i^\dagger \right]) \leq
\mathsf{SNR}$ for $1\le i\le L$.

The feedback path to the transmitter in an orthogonal frequency band is given by
\begin{equation}
\mathsf{R} \rightarrow \mathsf{T}_i: Y_{f,i} = H_{f,i} X_{f,i} + W_f,
\end{equation}
where $H_{f,i}$ is the MIMO fading channel for the feedback link to $i^{th}$ user, normalized much like the forward link. The feedback transmissions are also assumed to be power-limited, that is the reverse link has a power budget of $\frac{1}{T_{\rm coh}}{\rm trace}({\mathbb E}\left[ X_{f,i}X_{f,i}^\dagger \right]) \leq
\mathsf{SNR}_f$. Without loss of generality, we will assume a \emph{symmetry in resources}, such that $\mathsf{SNR} = \mathsf{SNR}_f$.

Finally, we assume that the receiver computes the common feedback indix $\jr\in\{0,\cdots,K-1\}$, which is broadcast over the downlink and received at transmitter $i$ as  $\jt_i\in\{0,\cdots,K-1\}$.

%
%

\subsection{Diversity-Multiplexing Tradeoff}

The diversity multiplexing tradeoff for single user MIMO channels was described in Section II.C. Here, we extend that discussion to MIMO MAC channels.
As before, we concentrate on single rate transmission. The dependence of rates on the
$\mathsf{SNR}$s is explicitly given by $R_s = r_s \log \mathsf{SNR}_s$.
We refer to ${\mathbf r}\triangleq (r_s)_{1\le s \le L}$ as the
multiplexing gains. Let ${\mathbf H}=\{H_1,\cdots,H_L\}$. Further, let the channel estimates at the receiver be $\widehat{\mathbf H}=\{\widehat{H}_1,\cdots,\widehat{H}_L\}$.

In a multiple access channel,
corresponding outage event is defined as the union over the events that the channel
cannot support target data rate for some subset of the users~\cite{tse04}, union over all the subsets. Hence, for a multiple access channel with $L$ users, each equipped
with $m$ transmit antennas, and a receiver with $n$ receive
antennas, the outage event is ${\cal O} \triangleq \mathop \bigcup \limits_S
{\cal O}_{S}$. The union is
taken over all subsets $S \subseteq \{1,2, \cdots, L\}$, and
${\cal O}_{S}$ is the set of all the channels where the
sum transmitted rate by these $|S|$ users is less than the maximum supportable rate by the MIMO link from these $|S|$ users to the destination. The
system is said to have diversity order of $d$ if
$\Pi({\cal O}) \doteq \mathsf{SNR}^{-d}$. The diversity multiplexing tradeoff for
the multiple access channel can be described as follows: given the multiplexing gains
${\mathbf r}$ for all the users, the diversity order that  can be
achieved describes the diversity-multiplexing tradeoff region.

The probability of outage with rate ${\mathbf R}=(R_1,R_2, \cdots,
R_L)$, transmit power $P(\jt_i)=P_i \forall \jt_i$ and perfect channel state information ${\mathbf H}$ at the receiver is
denoted by $\mathsf{S}({\mathbf R},{\mathbf P})\triangleq \Pi\left(\mathop \bigcup \limits_S
{\cal O}_S({\mathbf R}, {\mathbf P})\right)$. If we assume that the receiver knows the channel perfectly as ${\mathbf H}$, we denote the event $\mathop \cup \limits_S
{\cal O}_S({\mathbf R}, {\mathbf P}))$ by $U_{{\mathbf H}}({\mathbf R}, {\mathbf P})$ where we assume that ${\mathbf H}$ is the perfect channel knowledge at the receiver.

Let $D({\mathbf
r},{\mathbf p})$ be defined as $\mathsf{S}({\mathbf R},{\mathbf P}) \buildrel.\over= \mathsf{SNR}^{-D({\mathbf
r},{\mathbf p})}$ where ${\mathbf r}=(r_1,r_2, \cdots, r_L)$ and ${\mathbf
p}=(p_1,p_2, \cdots, p_L)$. We further denote function $G(r,p)$ in Section II by $G_{m,n}(r,p)$ to explicitly depict that this is for $m$ transmit and $n$ receive antennas.

\begin{lemma}\label{outlemma}\cite{vaneetno}
Let $p_s=p$ for all $1\le s \le
L$.  Also, let $\mathop \sum \limits_{i \in S}r_i\le \min(|S|m,n)$ for
all non-empty subsets $S$ of $\{1, 2, \cdots, L\}$. Then, \vspace{-.1in}
\begin{eqnarray}
D({\mathbf r}, {\mathbf p}) = \mathop {\min}
\limits_{S}G_{|S|m,n}\left(\mathop \sum \limits_{i \in S}r_i,\mathop
p\right).
\end{eqnarray}
\end{lemma}

\subsection{CSI$\widehat{\text{R}}\widehat{\text{T}}_q$: Estimated CSIR with Noisy Quantized Feedback}

All the results related to quantized feedback in this paper can be extended to multiple access channel where there is a feedback level sent from the receiver and all the transmitters receive this signal and adjust the power accordingly. To demonstrate the extension, we consider a symmetric system where all transmitters have a statistically identical channel to the receiver with identical average $\mathsf{SNR}$ and employ the same power control thresholds. Furthermore, we will only consider a single-bit feedback, which implies that simultaneously all transmitters will be instructed to use the low power level or the high power level. However, since we assume independent errors in the feedback links, each transmitter may or may not transmit at the right power level. Under the above conditions, an achievable diversity-multiplexing tradeoff is given by
\begin{theorem}
For $K=2$ and ${\mathbf r}=(r_1,\cdots, r_L)$ with $\mathop \sum \limits_{i \in S}r_i\le \min(|S|m,n)$ for
all non-empty subsets $S$ of $\{1, 2, \cdots, L\}$, an achievable diversity-multiplexing tradeoff for a multiple-access channel is given by ${\overline d}_{\widehat{\text{R}}\widehat{\text{T}}_q}({\mathbf r},2)=D({\mathbf r},{\mathbf 1}(1+D({\mathbf r},{\mathbf 1})))$ where ${\mathbf 1}x$ denotes $(x,x,\cdots,x)$.
\label{th:main-result-mac}
\end{theorem}
\begin{proof}
We will provide the main steps to prove the above result based on the following three-phase protocol.
First define $p_0=0$ and $p_j=D({\bf r},{\bf 1}(1+p_{j-1}))\forall j\ge 1$.

\noindent  {\bf Phase 1:} Each transmitter trains the receiver using power $\mathsf{SNR}$ to get the channel estimate $\widehat{H}_i$ at the receiver. On the basis of this training, the receiver decides a feedback level $\jr$ in the following way. We consider two power levels $P_i\doteq \mathsf{SNR}^{1+p_i}, i\in[0,1]$. We will state the exact constants for power control while proving that the average power constraint will be satisfied. The feedback index is
\[
\jr=\left\{ \begin{array}{l}
  1 \text{ if } U_{\widehat{\mathbf H}}({\mathbf R}+{\mathbf 1}\epsilon\log(\snr), {\mathbf 1}P_0)=1\\
  0  \text{ otherwise } \\
 \end{array}. \right.
 \]
Intuitively, we choose the higher of the two power levels if the lower power level is not sufficient to avoid outage (even for one of the users) based on the estimated channel.

\noindent  {\bf Phase 2:}  A feedback level $\jr\in\{0,1\}$ is sent from the receiver but each transmitter receives $\jt_i\in\{0,1\}$ according to the power controlled feedback scheme in Section \ref{sec:power-controlled feedback}.  Since the feedback links have i.i.d. errors, different transmitters may receive different feedback indices.

\noindent  {\bf Phase 3:}  The transmitter $k$ gets a feedback power level $\jt_k$, sends a training signal to the receiver at power level $P_{\jt_k}$ followed by data at power level $P_{\jt_k}$. The channel estimate based on this power-controlled training is denoted $\widehat{H}_{2,k}$. Let $\widetilde{H}_k=H_k-\widehat{H}_{2,k}$. Further, ${\widehat{\mathbf{H}}}_2=\{\widehat{H}_{2,1},\cdots,\widehat{H}_{2,L}\}$. Also, denote $\mathbf{H}_S$ as the $|S|m\times n$ matrix formed by concatenation of $H_i$ in $S$. Similarly, define ${\widehat{\mathbf{H}}_{2S}}$ and $\widehat{\mathbf H}_S$.

The outage probability is bounded from above by sum of outage probabilities for each transmitter. The analysis of Phase 3 is similar to Appendix \ref{phase3anal} since the estimation error in the third phase can be neglected for diversity multiplexing tradeoff purposes.
Now define $J$ as
 \[{J}=\left\{ \begin{array}{l}
  1 \text{ if } U_{{\widehat{\mathbf{H}}}_2}({\mathbf R}, {\mathbf 1}P_0)=1 \text{ and } U_{{\widehat{\mathbf{H}}}_2}({\mathbf R}, {\mathbf 1}P_1)=0\\
  0  \text{ otherwise } \\
 \end{array}. \right.\]
Since the third phase estimation error can be neglected, the outage probability is
\[
\Pi({\cal O})\doteq\Pi(U_{{\widehat{\mathbf{H}}}_2}({\mathbf R}, {\mathbf P}(\jt))=1).
\]
Hence, repeating the analysis of single user systems and using union bounds in Section \ref{mainsec}, we get the same results as in single user systems, but with $G$ replaced by $D$ and single multiplexing gain replaced by multiplexing gain vector.

We show by example how to extend all the steps. $\Pi(\jr=1,J=0)$ can be written as
\begin{eqnarray}
\label{eq:mac-outage}
&&\Pi(\jr=0,J=1)\nonumber\\
&=&\Pi\left(\log\det(I+\mathsf{SNR}\widehat{\mathbf H}_S\widehat{\mathbf H}^\dagger_S)\ge\sum_{i \in S}R_i \forall S  \text{ and }\right.\nonumber\\&& \quad \log\det(I+\mathsf{SNR}{{\widehat{\mathbf{H}}}_{2S}}{{\widehat{\mathbf{H}}}_{2S}}^\dagger)<\sum_{i \in S}R_i \text{ for some } S \nonumber\\&&\quad \left.\text{ and } \log\det(I+P_1{{\widehat{\mathbf{H}}}_{2S}}{{\widehat{\mathbf{H}}}_{2S}}^\dagger) \ge\sum_{i \in S}R_i \forall S\right)
\end{eqnarray}
We will now bound the above statement as follows. First define the following events
\begin{eqnarray}
A_S &=& \{\log\det(I+\mathsf{SNR}\widehat{\mathbf H}_S\widehat{\mathbf H}_S^\dagger)\ge\sum_{i \in S}R_i\}, \\
C_S &=& \{\log\det(I+\mathsf{SNR}{{\widehat{\mathbf{H}}}_{2S}}{{\widehat{\mathbf{H}}}_{2S}}^\dagger)<\sum_{i \in S}R_i\}, \\
D_S &=& \{\log\det(I+\mathsf{SNR}^{1+p_1}{{\widehat{\mathbf{H}}}_{2S}}{{\widehat{\mathbf{H}}}_{2S}}^\dagger)\ge\sum_{i \in S}R_i\}.
\end{eqnarray}

We note (\ref{eq:mac-outage}) is the probability of $\cap_SA_S\cap(\cup_SC_S)\cap\cap_SD_S\subseteq\cup_s(A_S\cap C_S)$. Hence,
\begin{eqnarray}&&\Pi(\jr=0,J=1)\nonumber\\
&\dot{\le}&
\sum_S\Pi\left(\log\det(I+\mathsf{SNR}\widehat{\mathbf H}_S\widehat{\mathbf H}_S^\dagger)\ge\sum_{i \in S}R_i \text{ and } \right.\nonumber\\&&\left.\quad \quad  \log\det(I+\mathsf{SNR}{{\widehat{\mathbf{H}}}_{2S}}{{\widehat{\mathbf{H}}}_{2S}}^\dagger)<\sum_{i \in S}R_i\right)
\end{eqnarray}
Note that this term is similar to that in single user, and for each $S$ is $\doteq 0$. Thus, the above probability $\doteq 0$.

Similarly, all other steps for single user MIMO channels can be extended to MIMO MAC system. Hence, the diversity order of $D({\mathbf r},{\mathbf 1}(1+D({\mathbf r},{\mathbf 1})))$ can be achieved with imperfect feedback and imperfect CSIR for $L$ transmitters.
\end{proof}
 In \cite{tse04}, diversity-multiplexing for multiple-access case was considered without feedback. The achievable diversity multiplexing without feedback is $D({\mathbf r},{\mathbf 1})$. In \cite{vaneet}, it was shown that with $1$ bit of perfect feedback, the diversity multiplexing of $D({\mathbf r},{\mathbf 1}(1+D({\mathbf r},{\mathbf 1})))$ can be achieved when the receiver knows perfect channel state information. In this paper, we show that the diversity multiplexing tradeoff of $D({\mathbf r},{\mathbf 1}(1+D({\mathbf r},{\mathbf 1})))$ can be achieved even when the receiver is trained on a noisy channel and the feedback index is also sent on an orthogonal noisy channel with power-controlled training and feedback.

\section{Conclusions}

In this paper, we find the diversity tradeoff for a non-symmetric FDD system in which the errors in MMSE channel estimate and the quantized feedback channel are accounted for a single user and a multiple access channel. We find that diversity multiplexing tradeoff of a system with $1$ bit of feedback over a noisy channel and MMSE channel estimate at the receiver is the same as that of a system with $1$ bit of perfect feedback and perfect channel estimate at the receiver. More importantly, we show that additional bits of feedback do not increase the diversity order of the system at constant rates.

The approach in this paper has also been used to improve the performance of a TDD system as is summarized in \cite{vaneetasil}. The two models, FDD and TDD, consider the two extreme cases of the correlations between the forward and the backward channel. As a next step, one can consider what happens if the forward and the feedback channel are correlated, but not exactly the same.

This paper suggests that one round of training provides a certain resolution to the channel gain which limits the diversity multiplexing tradeoff performance. The strategies can be extended to a multi-round communication between the sender and the receiver that allows better channel resolution at the nodes. This multi-round extension for both FDD and TDD models can be seen in \cite{advnew}.

Also, this paper assumes a Rayleigh fading channel model. The authors of \cite{genfade} consider a general model for fading which includes Rayleigh, Rician, Nakagami and Weibull distributions to find the diversity multiplexing tradeoff for a system with no feedback and perfect channel estimate at the receiver. The extension of the feedback cases to general fading models is still open.

Finally, the two way channel model can be extended to consider delays in the feedback channel. If there is a delay in the feedback process, the transmitter can decide to send some data as if there is no feedback till it receives feedback and then try to use the feedback to improve the diversity by sending power controlled data (possibly correlated with the data transmitted before feedback is received) in the remaining time.

\section{Acknowledgements}
The authors wish to thank Gajanana Krishna and Srikrishna Bhashyam for useful discussions related to this paper. We would also like to
thank the anonymous reviewers for many suggestions that improved this paper.


\appendices

\section{Proof of Theorem \ref{eigmimo}}\label{jointapp}
We first note some properties of $I_\nu(x)$, modified Bessel function of first kind, that will be used in the proof.
The series expansion of $I_\nu(x)$ is given as \cite[Equation 9.6.10]{handbook}\cite{bender03},
\begin{equation}\label{Ismall}
I_\nu(x)=\sum_{i=0}^\infty \frac{1}{i!(i+\nu)!}\left(\frac{x}{2}\right)^{2i+\nu}.
\end{equation}
When $|x|$ is large and $|\arg(x)|<\frac{\pi}{2}$ , asymptotic expansion of $I_\nu(x)$ is given by \cite[Equation 9.7.1]{handbook}
\begin{eqnarray}\label{Ilarge}
I_\nu(x)&\doteq& \frac{e^x}{\sqrt{2 \pi x}}\left\{1-\frac{\mu-1}{x}+\frac{(\mu-1)(\mu-9)}{2!(8x)^2}\right.\nonumber\\
&&\left.\quad -\frac{(\mu-1)(\mu-9)(\mu-25)}{3!(8x)^3}+\cdots\right\},
\end{eqnarray}
where $\mu=4\nu^2$.

Now, for the proof of Theorem \ref{eigmimo}, we will use Lemma \ref{si}. We will further suppose that $m_N=m$ without loss of generality. Let $A=\{i:\alpha_i+\widehat{\alpha}_i\le2\}$. 
We will evaluate $p({\ba,\bah})$ in the following five disjoint cases which comprise the whole space of possibilities.
\begin{enumerate}
\item $\min(\alpha_m,\widehat{\alpha}_m)\ge 1$ ( or $(\ba,\bah)\in E_m$).
\item $\min(\alpha_m,\widehat{\alpha}_m)< 1$, $\alpha_m+\widehat{\alpha}_m\ge2$.
\item $\min(\alpha_m,\widehat{\alpha}_m)< 1$, $\alpha_m+\widehat{\alpha}_m<2$, $\alpha_i \ne\widehat{\alpha}_i$ for some $i\in A$.
\item $\min(\alpha_m,\widehat{\alpha}_m)< 1$, $\alpha_m+\widehat{\alpha}_m<2$, $\alpha_i =\widehat{\alpha}_i$ for all $i\in A$, $(\ba,\bah)\notin \bigcup_{k=0}^{m-1}(E_{k})$.
\item $(\ba,\bah)\in \bigcup_{k=0}^{m-1}(E_{k})$.
\end{enumerate}

Now, we consider all of the cases one by one as follows.

\begin{enumerate}
\item $\min(\alpha_m,\widehat{\alpha}_m)\ge 1$ :
Using Equation \eqref{joint},
\begin{eqnarray}
p(\lambda,\widehat{\lambda})&=&\frac{\exp\left(-\frac{\sum_{k=1}^m \lambda_k+\widehat{\lambda}_k}{1-|\rho|^2}\right)\triangle(\lambda)\triangle(\widehat{\lambda})}{m!m!\Pi_{j=0}^{m-1}j!(j+\nu)!|\rho|^{mn-m}(1-|\rho|^2)^m}\nonumber\\
&&\quad \Pi_{k=1}^m (\sqrt{\lambda_k\widehat{\lambda}_k})^\nu\det\left|I_\nu\left(\frac{2|\rho|\sqrt{\lambda_k \widehat{\lambda}_l}}{1-|\rho|^2}\right)\right|\nonumber
\end{eqnarray}
Since $\rho=\frac{1}{\sqrt{1+\frac{m}{\mathsf{SNR}}}}$, we substitute $\rho\doteq1$ and $1-\rho^2\doteq 1/\mathsf{SNR}$ to obtain
\begin{eqnarray}\label{usep2}
p(\alpha,\widehat{\alpha})&\doteq&\exp\left(-\frac{\sum_{i=1}^m\mathsf{SNR}^{-\alpha_i}+\mathsf{SNR}^{-\widehat{\alpha}_i}}{1/\mathsf{SNR}}\right)\nonumber\\
&&\triangle(\mathsf{SNR}^{-\alpha})\triangle(\mathsf{SNR}^{-\widehat{\alpha}})\mathsf{SNR}^{-\sum_{i=1}^m(\alpha_i + \widehat{\alpha}_i)}\nonumber\\
&&(\sqrt{\mathsf{SNR}^{-\sum_{i=1}^m\alpha_i}\mathsf{SNR}^{-\sum_{i=1}^m\widehat{\alpha}_i}})^{(n-m)}\nonumber\\
&&\frac{ \det\left|I_{n-m}\left(\frac{2\sqrt{\mathsf{SNR}^{-\alpha_k} \mathsf{SNR}^{-\widehat{\alpha_l}}}}{1/\mathsf{SNR}}\right)\right|}{m!m!\Pi_{j=0}^{m-1}j!(j+n-m)!\frac{1}{\mathsf{SNR}^m}}.
\end{eqnarray}
%
%

As $\triangle(\mathsf{SNR}^{-\alpha})\doteq \mathsf{SNR}^{-\sum_{i=1}^m(i-1)\alpha_i}$, we get
\begin{eqnarray}\label{pen}
p(\alpha,\widehat{\alpha})&\doteq&\mathsf{SNR}^m \mathsf{SNR}^{-\sum_{i=1}^m (i+\frac{n-m}{2})(\alpha_i+\widehat{\alpha}_i)}\nonumber\\
&&\quad \det\left|I_{n-m}\left(\mathsf{SNR}^{1-\frac{\alpha_k+\widehat{\alpha}_l}{2}}\right)\right|.
\end{eqnarray}

We will now find $\det\left|I_{\nu}\left(\mathsf{SNR}^{1-\frac{\alpha_k+\widehat{\alpha}_l}{2}}\right)\right|$. Using \eqref{Ismall}, we get

\begin{eqnarray}
&&I_{\nu}\left(\mathsf{SNR}^{1-\frac{\alpha_k+\widehat{\alpha}_l}{2}}\right)\nonumber\\
&=&\sum_{i=0}^\infty\frac{1}{i! (i+\nu)!}\left(\frac{\mathsf{SNR}^{1-\frac{\alpha_k+\widehat{\alpha}_l}{2}}}{2}\right)^{2i+\nu}\nonumber\\
&=&\left(\frac{\mathsf{SNR}^{1-\frac{\alpha_k+\widehat{\alpha}_l}{2}}}{2}\right)^{\nu}\nonumber\\
&&\quad \sum_{i=0}^\infty\frac{1}{i! (i+\nu)!}\left(\frac{\mathsf{SNR}^{1-\frac{\alpha_k+\widehat{\alpha}_l}{2}}}{2}\right)^{2i}
\end{eqnarray}
Let $\mathsf{per}(k_1,k_2, \cdots, k_m)$ for $(k_1,k_2, \cdots, k_m)$ a permutation of $(1,\cdots,m)$ be defined as follows
\begin{eqnarray}
&& \mathsf{per}(k_1,k_2, \cdots, k_m)\nonumber\\
&\triangleq&\begin{cases}
  0  &\text{ if } (k_1,k_2, \cdots, k_m) \text{ is an even} \\
  & \quad \text{ permutation of $(1,\cdots,m)$} \\
  1  &\text{ if } (k_1,k_2, \cdots, k_m) \text{ is an odd}\\
  & \quad \text{ permutation of $(1,\cdots,m)$} \\
 \end{cases}.
\end{eqnarray}
Thus,
\begin{eqnarray}
&&\det\left|I_{\nu}\left(\mathsf{SNR}^{1-\frac{\alpha_k+\widehat{\alpha}_l}{2}}\right)\right|\nonumber\\&=&\sum_{\mathbf k}(-1)^{\mathsf{per}(k_1,k_2, \cdots, k_m)}\Pi_{l=1}^m\left(\frac{\mathsf{SNR}^{1-\frac{\alpha_{k_l}+\widehat{\alpha}_l}{2}}}{2}\right)^{\nu}\nonumber\\
&&\quad\sum_{i=0}^\infty\frac{1}{i! (i+\nu)!}\left(\frac{\mathsf{SNR}^{1-\frac{\alpha_{k_l}+\widehat{\alpha}_l}{2}}}{2}\right)^{2i}\nonumber\\
&\doteq&\mathsf{SNR}^{\nu\sum_{i=1}^m(1-\frac{\alpha_{i}+\widehat{\alpha}_i}{2})}\sum_{\mathbf k}(-1)^{per(k_1,k_2, \cdots, k_m)}\nonumber\\
&&\quad \Pi_{l=1}^m\sum_{i=0}^\infty\frac{1}{i! (i+\nu)!}\left(\frac{\mathsf{SNR}^{2-\alpha_{k_l}+\widehat{\alpha}_l}}{4}\right)^{i}
\end{eqnarray}

The above equation is same as Equation (56) in \cite{shin07} with $K=\frac{\sqrt{2}-1}{2}$, $\phi_i = \mathsf{SNR}^{1-\alpha_i}$ and $\lambda_i = \mathsf{SNR}^{1-\widehat{\alpha}_i}$, and hence
\begin{eqnarray}
&&\det\left|I_{\nu}\left(\mathsf{SNR}^{1-\frac{\alpha_k+\widehat{\alpha}_l}{2}}\right)\right|\nonumber\\
&\doteq&\mathsf{SNR}^{\nu\sum_{i=1}^m(1-\frac{\alpha_{i}+\widehat{\alpha}_i}{2})}\triangle(\mathsf{SNR}^{1-\alpha_i})\triangle(\mathsf{SNR}^{1-\widehat{\alpha}_i})\nonumber\\
&\doteq&\mathsf{SNR}^{\nu\sum_{i=1}^m(1-\frac{\alpha_{i}+\widehat{\alpha}_i}{2})}\mathsf{SNR}^{2\sum_{i=1}^m(i-1)(1-\frac{\alpha_{i}+\widehat{\alpha}_i}{2})}\nonumber\\
&\doteq&\mathsf{SNR}^{\sum_{i=1}^m(2i-2+\nu)\left(1-\frac{\alpha_{i}+\widehat{\alpha}_i}{2}\right)}.
\end{eqnarray}

Substituting in Equation \eqref{pen}, we get
\begin{eqnarray}
p(\alpha,\widehat{\alpha})&\doteq&\mathsf{SNR}^m \mathsf{SNR}^{-\sum_{i=1}^m (i+\frac{n-m}{2})(\alpha_i+\widehat{\alpha}_i)}\nonumber\\
&& \quad \mathsf{SNR}^{\sum_{i=1}^m(2i-2+\nu)\left(1-\frac{\alpha_{i}+\widehat{\alpha}_i}{2}\right)}
\nonumber\\&\doteq&\mathsf{SNR}^{mn} \mathsf{SNR}^{-\sum_{i=1}^m (i+\frac{n-m}{2})(\alpha_i+\widehat{\alpha}_i)}\nonumber\\
&&\quad \mathsf{SNR}^{-\sum_{i=1}^m(i-1+\frac{n-m}{2})(\alpha_{i}+\widehat{\alpha}_i)}
\nonumber\\&\doteq&\mathsf{SNR}^{mn} \mathsf{SNR}^{-\sum_{i=1}^m (2i+n-m-1)(\alpha_i+\widehat{\alpha}_i)}\nonumber\\
&=&e_m.
\end{eqnarray}

\item $\min(\alpha_m,\widehat{\alpha}_m)< 1$, $\alpha_m+\widehat{\alpha}_m\ge2$: Using Equation \eqref{usep2}, we see that all terms except $\exp\left(-\frac{\sum_{i=1}^m\mathsf{SNR}^{-\alpha_i}+\mathsf{SNR}^{-\widehat{\alpha}_i}}{1/\mathsf{SNR}}\right)$ remain the same and hence are polynomial in $\mathsf{SNR}$ while $\exp\left(-\frac{\sum_{i=1}^m\mathsf{SNR}^{-\alpha_i}+\mathsf{SNR}^{-\widehat{\alpha}_i}}{1/\mathsf{SNR}}\right)$ decreases exponentially in $\mathsf{SNR}$ and hence $p(\alpha,\widehat{\alpha})\doteq0$.

\item $\min(\alpha_m,\widehat{\alpha}_m)< 1$, $\alpha_m+\widehat{\alpha}_m<2$, $\alpha_i \ne\widehat{\alpha}_i$ for some $i\in A$: We will prove $p(\alpha,\widehat{\alpha})\doteq0$ in this case. For this we consider $\exp\left(-\frac{\sum_{k=1}^m \lambda_k+\widehat{\lambda}_k}{1-|\rho|^2}\right)\det\left|I_\nu\left(\frac{2|\rho|\sqrt{\lambda_k \widehat{\lambda}_l}}{1-|\rho|^2}\right)\right|$ and prove this part to decrease exponentially and we would be done since rest of the terms are polynomial in $\mathsf{SNR}$. Using \eqref{Ilarge},
\begin{eqnarray}\label{chn}
&&\det\left|I_\nu\left(\frac{2|\rho|\sqrt{\lambda_k \widehat{\lambda}_l}}{1-|\rho|^2}\right)\right|\nonumber\\
&\doteq& \left(\sum_{\mathbf k}(-1)^{\mathsf{per}(k_1,k_2, \cdots, k_m)}\Pi_{l=1, \alpha_{k_l}+\widehat{\alpha}_l\le 2}^m\right.\nonumber\\
&&\quad \frac{\exp\left(\frac{2|\rho|\sqrt{\lambda_{k_l}\widehat{\lambda}_l}}{1-\rho^2}\right)}
{\sqrt{\frac{2\sqrt{\lambda_{k_l} \widehat{\lambda}_l}}{1-|\rho|^2}}}(1-\frac{\mu-1}{\frac{16\sqrt{\lambda_{k_l}\widehat{\lambda}_l}}{1-\rho^2}}+\cdots)\nonumber\\
&&\quad \left.\Pi_{l=1, \alpha_{k_l}+\widehat{\alpha}_l> 2}^m \mathsf{Poly}(\mathsf{SNR})\right),
\end{eqnarray}
where $\mathsf{Poly}(\mathsf{SNR})$ represents the term that are polynomial in $\mathsf{SNR}$. First observe that
\begin{equation}
\exp\left(-\frac{\lambda_{k_l}+\widehat{\lambda}_l}{1-\rho^2}\right)\exp\left(\frac{2|\rho|\sqrt{\lambda_{k_l}\widehat{\lambda}_l}}{1-\rho^2}\right)\le 1,\end{equation} and hence the product of such terms cannot increase exponentially with $\mathsf{SNR}$.

\begin{eqnarray}
&&\exp\left(-\frac{\sum_{k=1}^m \lambda_k+\widehat{\lambda}_k}{1-|\rho|^2}\right)\det\left|I_\nu\left(\frac{2|\rho|\sqrt{\lambda_k \widehat{\lambda}_l}}{1-|\rho|^2}\right)\right|\nonumber\\
&\doteq&\sum_{\mathbf k}(-1)^{\mathsf{per}(k_1,k_2, \cdots, k_m)}\nonumber\\
&&\quad \Pi_{l=1}^m \exp(-\mathsf{SNR}^{1-\min(\alpha_{k_l},\widehat{\alpha}_l)})\nonumber\\
&&\quad \Pi_{l=1, \alpha_{k_l}+\widehat{\alpha}_l\le 2}^m{\exp(\mathsf{SNR}^{1-\frac{\alpha_{k_l}+\widehat{\alpha}_l}{2}})}
 \mathsf{Poly}(\mathsf{SNR})\nonumber\\
 &\doteq& \sum_{\mathbf k}(-1)^{\mathsf{per}(k_1,k_2, \cdots, k_m)}\nonumber\\
 &&\quad \Pi_{l=1,\alpha_{k_l}+\widehat{\alpha}_l>2}^m \exp(-\mathsf{SNR}^{1-\min(\alpha_{k_l},\widehat{\alpha}_l)})\nonumber\\
 &&\quad \Pi_{l=1, \alpha_{k_l}+\widehat{\alpha}_l\le 2}^m \left(\exp(-\mathsf{SNR}^{1-\min(\alpha_{k_l},\widehat{\alpha}_l)})\right.\nonumber\\
 &&\quad \left. \exp(\mathsf{SNR}^{1-\frac{\alpha_{k_l}+\widehat{\alpha}_l}{2}})\right) \mathsf{Poly}(\mathsf{SNR})\nonumber\\
 &\dot{\le}& \sum_{\mathbf k}(-1)^{\mathsf{per}(k_1,k_2, \cdots, k_m)}\nonumber\\
 &&\quad \Pi_{l=1,\alpha_{k_l}+\widehat{\alpha}_l>2}^m {\mathbf 1}_{\min(\alpha_{k_l},\widehat{\alpha}_l)\ge1}\nonumber\\
  &&\quad \Pi_{l=1, \alpha_{k_l}+\widehat{\alpha}_l\le 2}^m {\mathbf 1}_{\alpha_{k_l}=\widehat{\alpha}_l} \mathsf{Poly}(\mathsf{SNR})
\end{eqnarray}
Now, we will show that each term under the sum decays exponentially with $\mathsf{SNR}$. For this, the product of the above indicators should be $0$. Let us now consider all the scenarios when the above product of indicators is $1$. If some $\widehat{\alpha}_l< 1$, and pairs with $\alpha_{k_l}$ to give $\widehat{\alpha}_l+\alpha_{k_l}\le 2$, then the two must be equal and if it pairs to give $\widehat{\alpha}_l+\alpha_{k_l}> 2$, the product of indicators will always be $0$. If $\widehat{\alpha}_l\ge 1$ and pairs with $\alpha_{k_l}$ to give $\widehat{\alpha}_l+\alpha_{k_l}\le 2$, it can only happen when the two are equal and if it pairs to give $\widehat{\alpha}_l+\alpha_{k_l}> 2$, then $\alpha_{k_l}\ge 1$. Hence the only pairing that will work is that all elements of $\alpha_i<1$ are matched to $\widehat{\alpha_j}<1$ and also $\alpha_i\ge1$ is not mapped to $\widehat{\alpha}_i<1$. This can happen only when $\alpha_i=\widehat{\alpha}_i$ whenever $\alpha_i<1$ or $\widehat{\alpha}_i<1$ and all the rest are $\ge 1$. If $\alpha_i\ne \widehat{\alpha}_i$ for some $i\in A$, the above condition do not hold. This proves Case 3.

\item $\min(\alpha_m,\widehat{\alpha}_m)< 1$, $\alpha_m+\widehat{\alpha}_m<2$, $\alpha_i =\widehat{\alpha}_i$ for all $i\in A$, $(\ba,\bah)\notin \bigcup_{k=0}^{m-1}(E_{k})$: We see that all the analysis of Case 3 holds for this case and hence if $\alpha_i =\widehat{\alpha}_i$ for all $i\in A$ and $(\alpha,\widehat{\alpha})\notin \bigcup_{k=0}^{m-1}(E_{k})$, an element $\ge 1$ is mapped to an element $<1$ which makes the product of indicators zero and hence the probability decreases exponentially with $\mathsf{SNR}$.

\item $\ba,\bah\in \bigcup_{k=0}^{m-1}(E_{k})$. These are the cases we sum over in the statement of the Lemma. In each of these cases, $p(\alpha,\widehat{\alpha})$ exist. When we integrate over $\widehat{\alpha}$, we find that integral of $\sum_{k}e_{k}{\mathbf 1}_{E_{k}}$ w.r.t. $\widehat{\alpha}$ is $\Pi_{i=1}^m\mathsf{SNR}^{-(2i-1+n-m)\alpha_i}{\mathbf 1}_{\alpha_{1}, \cdots ,\alpha_k\ge1}{\mathbf 1}_{0\le\alpha_{k+1}, \cdots ,\alpha_{m_N}<1}$. We next note that in order to find the polynomial expressions associated, the two exponentials multiplication cannot decrease with $\mathsf{SNR}$ for which we would need $k_l=l$ in \eqref{chn} (since all the $\lambda$'s and $\widehat{\lambda}$'s are ordered) and hence from all the expressions, we find that $p(\ba,\bah)=p_1(\ba)p_2(\bah)$ for any $E_k$. Using the separability, we find that
    \begin{eqnarray}
    e_k&=&\mathsf{SNR}^{k(n-m+k)}\Pi_{i=1}^{k}\mathsf{SNR}^{-(2i-1+|n-m|)\widehat\alpha_i}\nonumber\\
    && \quad \Pi_{i=1}^{m_N}\mathsf{SNR}^{-(2i-1+|n-m|)\alpha_i}.
    \end{eqnarray}
\end{enumerate}

\section{Proof of Theorem \ref{thrmif}}\label{apdxif}

In this appendix, we prove that the diversity in the statement of Theorem \ref{thrmif} can be achieved.
Let $\mathsf{S}(R,P)$ be the probability of outage when the transmitter uses power level $P$ and rate $R$ is required. If $P\doteq \mathsf{SNR}^p$, then $\mathsf{S}(R,P)\doteq \mathsf{SNR}^{-G(r,p)}$ and
\[
\Pi({\cal O}) \dot{\le} \mathsf{S}(R,P_{K-1})+\sum_{i=1}^{K-1}\Pi(\jt<i,\jr=i).
\]
We assign the feedback index $\jr$ at the receiver as
\[
\jr =
\begin{cases}
\arg \min_{i \in \mathbb{I}}  i,  & \mathbb{I} = \{ k : \log \det(I+{H}{H}^\dagger {P_{k}})\ge R,\\
& \quad k \in \{0,\cdots,K-1\} \}  \\
  K-1,  & \text{ if the set $\mathbb{I}$ is empty } \\
\end{cases}.
\]
We will now find $\Pi(\jt=i)$ for $i\ge 1$ as
\begin{eqnarray}
&&\Pi(\jt=i)\nonumber\\
&=&\sum_{j=0}^{K-1}{\Pi(\jt=i,\jr=j)}\nonumber\\
&=&\sum_{j=0}^{K-1}{\Pi(\jt=i|\jr=j)\Pi(\jr=j)}\nonumber\\
&\doteq&\Pi(\jr=i)+\sum_{j=i+1}^{K-1}{\Pi(\jt=i|\jr=j)\Pi(\jr=j)}\nonumber\\
&\doteq&\Pi(\jr=i).
\end{eqnarray}
The above steps follow from the fact that $\Pi(\jt=i|\jr=j)\doteq0$ if $j<i$.
Further,
\begin{eqnarray}
&&\sum_{j=i+1}^{K-1}{\Pi(\jt=i|\jr=j)\Pi(\jr=j)}\nonumber\\&\dot{\le}&\sum_{j=i+1}^{K-1}{\Pi(\jr=j)}\dot{\le}\Pi(\jr=i).\end{eqnarray}
Let the power levels be chosen as
\[
 P_i  = \left\{ \begin{array}{l}
 \frac{\mathsf{SNR}}{K}\text{ when $i = 0$} \\
 \frac{\mathsf{SNR}}{{K\mathsf{S}(R ,P_{i-1})}}\text{ when $i > 0$} \\
 \end{array} .
 \right.
\]
Further, $Q_i\le P_i$.
We first note that the power constraints are satisfied. Note that as before, $P_i=\mathsf{SNR}^{1+p_i}$ and $Q_i=\mathsf{SNR}^{q_i}$. Also, $P_i\doteq \mathsf{SNR}^{1+d_{\text{RT}_q}(r,i)} \forall i$ and $\mathsf{S}(R,P_i)\doteq \mathsf{SNR}^{-d_{\text{RT}_q}(r,i+1)}$. Thus, $\Pi(\jr=i)=\Pi(\jt=i)\doteq \mathsf{SNR}^{-d_{\text{RT}_q}(r,i)}$.

We will now show that using $P_i$ as given above, we achieve the desired diversity multiplexing tradeoff using the following computation.

\begin{eqnarray}
&&\Pi({\cal O})\nonumber\\
&\dot{\le}& \mathsf{S}(R,P_{K-1})+\sum_{i=1}^{K-1}\Pi(\jt<i,\jr=i)\nonumber\\
&\doteq& \mathsf{SNR}^{-d_{\text{RT}_q}(r,K)}+\sum_{i=1}^{K-1}\Pi(\jt<i|\jr=i)\Pi(\jr=i)\nonumber\\
&\doteq& \mathsf{SNR}^{-d_{\text{RT}_q}(r,K)}+\sum_{i=1}^{K-1}\Pi(\jt<i|\jr=i)\mathsf{SNR}^{-d_{\text{RT}_q}(r,i)}\nonumber\\
&\doteq& \mathsf{SNR}^{-d_{\text{RT}_q}(r,K)}\nonumber\\
&& \quad +\sum_{i=1}^{K-1}\mathsf{SNR}^{-mn(\max(q_i,0)-\max(q_{i-1},0))}\mathsf{SNR}^{-d_{\text{RT}_q}(r,i)}\nonumber\\
&\doteq& \mathsf{SNR}^{-{\overline d}_{\text{R}{\widehat{\text{T}}}_q}(r,K)}.
\end{eqnarray}

\section{Proof of Theorem \ref{thrm:csihatrt}}\label{app:csihatrt}
We will first show that the diversity cannot be greater than $mn(mn+1)$, and then show that diversity of $G(r,1+G(r,1)$ can be achieved with 1 bit of feedback.
\subsection{Converse}
In this subsection, we will prove that we cannot get more diversity with $>2$ levels of feedback with imperfect CSIR as compared to $1$ bit of feedback with perfect CSIR at zero multiplexing.

For this, we assume that the third phase is perfect and thus $\widetilde{H}=0$, and $\hr=H$. Thus,
\[
\Pi({\cal O})\doteq\Pi(\log \det(I+P(\jr)HH^\dagger)<R).
\]
Define $J$ as
\[
J =
\begin{cases}
\arg \min_{i \in \mathbb{I}}  i,  & \mathbb{I} = \{ k : \log \det(I+HH^\dagger {P_{k}})\ge R, \\
& \quad k \in \{0,\cdots,K-1\} \}  \\
  0,  & \text{ if the set $\mathbb{I}$ is empty } \\
\end{cases}.
\]
The probability of outage is then
\begin{eqnarray}\label{comppouta}
\Pi({\cal O})&\doteq&\Pi(\log \det(I+P(\jr)HH^\dagger)<R)\nonumber\\
&\dot{\ge}& \Pi(\log \det(I+P(\jr)HH^\dagger)<R,\jr=1)\nonumber\\
&\doteq& \Pi(\log \det(I+P_1HH^\dagger)<R,\nonumber\\
&& \quad \log \det(I+P_0\widehat{H}\widehat{H}^\dagger)<R+\epsilon\log(\snr),\nonumber\\&&\quad \log \det(I+P_1\widehat{H}\widehat{H}^\dagger)\ge \nonumber\\&&\quad R+\epsilon\log(\snr)).
\end{eqnarray}

Let $\lambda_j$ and $\widehat\lambda_j$ be the eigenvalues of $HH^\dagger$ and $\widehat{H}\widehat{H}^\dagger$ respectively. Further let $\alpha_j$ and $\widehat\alpha_j$ be the negative $\mathsf{SNR}$ exponents of the corresponding eigenvalues. Then, Equation \eqref{comppouta} reduces to
\begin{eqnarray}\label{comppout2}
\Pi({\cal O})&\dot{\ge}& \Pi(\log \det(I+P_1HH^\dagger)<R,\nonumber\\
&& \quad \log \det(I+P_0\widehat{H}\widehat{H}^\dagger)<R+\epsilon\log(\snr),\nonumber\\&& \quad \log \det(I+P_1\widehat{H}\widehat{H}^\dagger)\ge R+\epsilon\log(\snr))\nonumber\\
&\dot{\ge}&\Pi( \Pi_{j=1}^m(1+\lambda_j\mathsf{SNR}^{1+p_{1}})<\mathsf{SNR}^r \text{ and } \nonumber\\&& \quad \Pi_{j=1}^m(1+\widehat\lambda_j\mathsf{SNR})< \mathsf{SNR}^{r+\epsilon} \text{ and }\nonumber\\&& \quad \Pi_{j=1}^m(1+\widehat\lambda_j\mathsf{SNR}^{1+p_{1}})> \mathsf{SNR}^{r+\epsilon} )\nonumber\\
&\doteq&\Pi( \sum_{j=1}^m(1+p_{1}-\alpha_j)^+<r \text{ and }  \nonumber\\ &&\quad \sum_{j=1}^m(1-\widehat\alpha_j)^+< r+\epsilon \text{ and } \nonumber\\ &&\quad \sum_{j=1}^m(1+p_{1}-\widehat\alpha_j)^+> r+\epsilon).
\end{eqnarray}
For $r\to 0$, the above reduces to
\begin{eqnarray}\label{comppout2}
\Pi({\cal O})&\dot{\ge}& \Pi( \sum_{j=1}^m(1+p_1-\alpha_j)^+\le 0 \text{ and } \nonumber\\&& \quad \sum_{j=1}^m(1-\widehat\alpha_j)^+< \epsilon \text{ and } \nonumber\\&& \quad \sum_{j=1}^m(1+p_{1}-\widehat\alpha_j)^+> \epsilon).
\end{eqnarray}
Thus, the choice of $\alpha_j=1+p_1$ and all $\widehat\alpha_j=1$ for $\epsilon\to 0$ can be used for the outer bound on outage probability. This gives $\Pi({\cal O})\dot{\ge}\snr^{-mn(p_1+1)}$. As $p_1\le mn$, the diversity $\le mn(mn+1)$.

\subsection{Achievability}

In this subsection, we will show that the diversity gain of $G(r,1+G(r,1))$ can be achieved with $1$ bit of feedback with imperfect CSIR. We will first prove that the interference error due to $\widetilde{H}$ in third phase do not make a difference and thus can be removed.
\subsubsection{Analysis of Phase 3}\label{phase3anal}
Let the actual channel be $H$ while $\widetilde{H}$ be the error in the estimate of $H$ in the third phase of training which is power controlled. As a result, we obtain
\begin{eqnarray}
\Pi({\cal O})&=& \sum_{j=0}^{K-1}\Pi({\cal O},\jt=j)\nonumber\\
&\doteq&\sum_{j=0}^{K-1}\Pi(\log\det(1+\frac{P_j\hr\hr^\dagger}{1+P_j{\rm trace}(\widetilde{H}\widetilde{H}^\dagger)})<R,\nonumber\\&& \quad \jt=j)\nonumber .
\end{eqnarray}
Consider any term in the sum above, the receiver is trained with power $P_j$, and hence let eigen-values of $P_j\widetilde{H}\widetilde{H}^\dagger$ be $\widetilde{\lambda}_i\doteq \mathsf{SNR}^{-\widetilde{\alpha}_i}$. Let $\widetilde{\ba}=(\widetilde{\alpha}_1,\cdots,\widetilde{\alpha}_m)$ Then,
\begin{eqnarray}
&&\Pi({\cal O})\nonumber\\&\doteq&\sum_{j=0}^{K-1}\Pi(\log\det(1+\nonumber\\&&\quad \frac{P_j\hr\hr^\dagger}{1+P_j{\rm trace}(\widetilde{H}\widetilde{H}^\dagger)})<R,\jt=j)\nonumber\\
&\doteq&\sum_{j=0}^{K-1}\Pi(\log\det(1+{P_j\hr\hr^\dagger}\mathsf{SNR^{-(-\min\widetilde{\ba})^+}})<R,\nonumber\\&&\quad \jt=j).\nonumber
\end{eqnarray}
Now, since $\hr$ and $\widetilde{H}$ are uncorrelated, probability that $(-\min\widetilde{\ba})^+>0$ decreases higher than polynomial in $\mathsf{SNR}$ and hence,
\begin{eqnarray}
\Pi({\cal O})&\doteq&\sum_{j=0}^{K-1}\Pi(\log\det(1+{P_j\hr\hr^\dagger}\mathsf{SNR^{-(-\min\widetilde{\ba})^+}})\nonumber\\&&\quad <R,\jt=j)\nonumber\\
&\doteq&\sum_{j=0}^{K-1}\Pi(\log\det(1+{P_j\hr\hr^\dagger})<R,\jt=j)\nonumber\\
&\doteq&\Pi(\log\det(1+{P(\jt)\hr\hr^\dagger})<R).\nonumber\\
\end{eqnarray}

%

Thus, \[\Pi({\cal O})\doteq\Pi(\log \det(I+P(\jt)\hr\hr^\dagger)<R)\]

Note that the correlations between $\hr$ and $\widehat{H}$ are same as between $H$ and $\widehat{H}$ and thus, there is no difference in using $H$ in place of $\hr$
for the purpose of calculating the diversity multiplexing tradeoff. We will now only focus on $K=2$.

\subsubsection{Analysis of Phase 1}
In this section, we see how the feedback error in first phase decays  with $\mathsf{SNR}$.
\[\jr=\left\{ \begin{array}{l}
  1 \text{ if } \log \det(I+\widehat{H}\widehat{H}^\dagger {P_{0}})<R+\epsilon\log(\snr)\\
  0  \text{ if } \log \det(I+\widehat{H}\widehat{H}^\dagger {P_{0}})\ge R+\epsilon\log(\snr) \\
 \end{array} \right.\]
 Let us define $J$ as
\[{J}=\left\{ \begin{array}{l}
  1  \text{ if } \log \det(I+{\hr}{\hr}^\dagger P_{0})<R \text{ and }\\
  \quad \quad  \log \det(I+\hr\hr^\dagger P_1)\ge R\\
  0  \text{ otherwise } \\
 \end{array} .\right.\]

 The probability of outage is then
\begin{eqnarray}\label{comppout}
&&\Pi({\cal O})\nonumber\\
&\doteq&\Pi(\log \det(I+P(\jt)\hr\hr^\dagger)<R)\nonumber\\
&\doteq& \Pi(\log \det(I+P(\jt)\hr\hr^\dagger)<R,J=0,\jr=0)\nonumber\\
&&\quad + \Pi(\log \det(I+P(\jt)\hr\hr^\dagger)<R,\jr=1)\nonumber\\
&&\quad + \Pi(\log \det(I+P(\jt)\hr\hr^\dagger)<R,J=1,\jr=0)\nonumber\\
&\dot{\le}& \Pi(\log \det(I+P_{1}\hr\hr^\dagger)<R,\jr=0)\nonumber\\
&&\quad + \Pi(\log \det(I+P_{1}\hr\hr^\dagger)<R,\jr=1)\nonumber\\
&&\quad +\Pi(\jr=0,J=1)\nonumber\\
&\doteq& \Pi(\log \det(I+P_{1}\hr\hr^\dagger)<R)\nonumber\\&&\quad + \Pi(\jr=0,J=1).
\end{eqnarray}
Denote eigenvalues of $\widehat{H}\widehat{H}^\dagger$ by $\widehat\lambda_i$ and the negative $\mathsf{SNR}$ exponents of $\widehat\lambda_i$ as $\widehat\alpha_i$. Also denote eigenvalues of $\hr\hr^\dagger$ by $\lambda_i$ and the negative $\mathsf{SNR}$ exponents of $\lambda_i$ as $\alpha_i$.

Then, $\Pi(\jr=0,J=1)$ can be bounded as
\begin{eqnarray}
&&\Pi(\jr=0,J=1)\nonumber\\
&=&\Pi\left(\log \det(I+\hr\hr^\dagger \mathsf{SNR})<R \text{ and }\right. \nonumber\\ && \quad \log \det(I+\hr\hr^\dagger \mathsf{SNR}^{1+p_1})\ge R \text{ and }\nonumber\\&& \quad \left.\log \det(I+\widehat{H}\widehat{H}^\dagger \mathsf{SNR})\ge R+\epsilon\log(\snr) \right)\nonumber\\&\doteq&
\Pi\left( \Pi_{i=1}^m(1+\lambda_i\mathsf{SNR})<\mathsf{SNR}^r \text{ and } \right.\nonumber\\&& \quad  \Pi_{i=1}^m(1+\lambda_i\mathsf{SNR}^{1+p_1})\ge \mathsf{SNR}^r \text{ and }\nonumber\\&& \quad \left. \Pi_{i=1}^m(1+\widehat\lambda_i\mathsf{SNR})\ge \mathsf{SNR}^{r+\epsilon} \right)\nonumber\\
&\dot{\le}&\Pi\left( \sum_{i=1}^m(1-\alpha_i)^+\le r \text{ and } \right.\nonumber\\&& \quad \sum_{i=1}^m(1+p_1-\alpha_i)^+\ge r \text{ and }\nonumber\\&& \quad \left. \sum_{i=1}^m(1-\widehat\alpha_i)^+\ge r+\epsilon\right)\nonumber\\
&\doteq& 0.
\end{eqnarray}

The last step follows since expanding in terms of $E_k$ gives $\sum_{i=1}^m(1-\alpha_i)^+=\sum_{i=1}^m(1-\widehat{\alpha}_i)^+$ for all cases of non exponentially decreasing probability.

We now show that the diversity order of $G(r,1+G(r,1))$ can be achieved. From Equation \eqref{comppout}, the probability of outage for $K=2$ is
\begin{eqnarray}
\Pi({\cal O})&\dot{\le}& \Pi(\log \det(I+P_{1}\hr\hr^\dagger)<R)\nonumber\\
&& \quad + \Pi(\jr=0,J=1)\nonumber\\
&\dot{\le}& \mathsf{SNR}^{-G(r,1+p_1)}.
\end{eqnarray}
Thus, diversity order of $G(r,1+G(r,1))$ can be achieved which is the optimal diversity order for perfect training in \cite{kim07}.

Lastly, we show that the power constraint can be satisfied with the above choice of powers.

\begin{lemma} Power constraint is satisfied for $K=2$.
\end{lemma}
\begin{proof}
Let $P_0=\frac{\mathsf{SNR}}{2}$ and $P_1 =
\frac{\mathsf{SNR}}{2(\Pi(\jr=1))}$.

The power constraint is $P_0 \Pi(\jt=0)+P_1 \Pi(\jt=1)\le \mathsf{SNR}$, which trivially holds.
\end{proof}

\begin{biographynophoto}{Vaneet Aggarwal}
 received the B.Tech. degree in 2005 from the Indian Institute of Technology,
Kanpur, India and the M.A. degree in 2007 from Princeton University,
Princeton, NJ, USA, both in Electrical Engineering. He is currently
pursuing the Ph.D. degree in Electrical Engineering at Princeton
University, Princeton, NJ, USA.

His research interests are in applications of information and coding theory to wireless systems and quantum error correction. He was the recipient of Princeton University's Porter Ogden Jacobus Honorific Fellowship in 2009.
\end{biographynophoto}

\begin{biographynophoto}{Ashutosh Sabharwal}(S'91 - M'99 - SM'04) received the B.Tech. degree from
the Indian Institute of Technology, New Delhi, in 1993 and the M.S. and
Ph.D. degrees from The Ohio State University, Columbus, in 1995 and 1999,
respectively.

He is currently an Assistant Professor in the Department of Electrical and Computer Engineering and
also the Director of Center for Multimedia Communications at Rice University, Houston, TX. His research interests are in the areas of information theory and
communication algorithms for wireless systems.
Dr. Sabharwal was the recipient of Presidential Dissertation Fellowship
award in 1998.
\end{biographynophoto}

\end{document}